\documentclass[12pt, a4paper]{article}
\usepackage{bbold}
\usepackage{amssymb}
\usepackage{amsmath}
\usepackage{mathrsfs}
\usepackage{authblk}
\usepackage{float}
\usepackage{cite}
\usepackage{lipsum}
\usepackage{fullpage}
\usepackage{hyperref}
\usepackage{slashed}
\usepackage{graphicx}
\usepackage[titletoc, title]{appendix}
\usepackage{authblk}
\usepackage{bbm}
\usepackage{dsfont}
\title{\textbf{Fermions on the Worldsheet of Effective Strings via Coset Construction}}
\author{Ali Mohsen\footnote{ahm302@nyu.edu}}
\affil{\normalsize{\textit{Center of Cosmology and Particle Physics,} \\ \textit{Department of Physics, New York University} \\ \textit{New York, NY, 10003, USA}}}
\date{}
\begin{document}

\maketitle

\abstract{In this paper the detailed CCWZ procedure for introducing fermions on the world sheet of a string propagating in flat space-time is presented. The theory of nonlinear realizations is used to derive the transformation as well as the interactions of fermionic matter fields under arbitrary spinorial representations of the unbroken subgroup. This demonstrates that even for non-supersymmetric spinors, the interactions are still severely restricted by the nonlinearly realized symmetry. We also explain how supersymmetric models provide an example for this construction with Goldstinos as matter fields, and how one can use the $\kappa$-symmetry of the Green Schwarz action in particular, to verify this nonlinear transformation for a specific matter field representation. We finally restrict the target space dimension without reference to supersymmetry, but rather by imposing one-loop integrability on a fermionic string that nonlinearly realizes Poincare symmetry. This singles out the critical dimension $D=10$ for heterotic, GS and RNS supersymmetric strings.}
\newpage
\mbox{}
  \tableofcontents 
 
 \newpage 
 
 \section{Introduction}

One of the corner stones of modern particle physics is the discovery of hidden symmetries; those symmetries of the theory which are not realized in its spectrum and do not preserve the vacuum, and are said to be spontaneously broken. After it was shown that for each generator of such internal symmetry there exists a massless boson, the lightness of the pion was attributed to such a mechanism where the unbroken subgroup is the isospin group $SU(2)$, and where the broken component was to be experimentally deduced from processes with multi-pion emission using ``current algebra" methods. Subsequently a different and more intuitive technique based on effective field theory was introduced \cite{weinberg}, and generalized to the CCWZ procedure for arbitrary internal symmetry quotient groups \cite{CCWZ} and finally to spontaneously broken space-time symmetries \cite{salam,manohar, ccwz0}.

For broken space-time symmetries the naive counting of Goldstone modes fails, and the correct counting procedure, shows for example that physical Goldstone bosons for the breaking of the conformal group down to the Poincare group\\ $SO(2,D)/SO(1,D-1)$ only correspond to single dilatation mode and the remaining $D$ broken generators don't give rise to dynamical fields. In this paper we will focus on the long wavelength physics of confining strings in Poincare invariant theories. The presence of such long string background breaks down the full Poincare group. That is because now only boosts along the string and rotations around it preserve such vacuum, whereas transverse translations and boosts, as well as rotations around axes orthogonal to the string axis don't. Therefor we must consider the coset construction of $ISO(1,D-1)/ISO(1,1)\times SO(D-2)$, where the only surviving symmetries are the $ISO(1,1)$ world sheet Poincare invariance, and $SO(D-2)$ which represent rotations around the string axis. Here Goldstone bosons only correspond to the $D-2$ broken translational generators which describe transverse low energy excitations, whereas the remaining broken $2(D-2)$ rotations will correspond to auxiliary fields (section \ref{1}), which will eventually be expressed in terms of the physical fields (section \ref{projecting}).

This procedure has been successfully implemented for both the p-brane and the super p-brane \cite{west1}, where the reparametrization invariance was left unfixed. In this paper we work in static gauge \cite{cooper}, suitable for effective string calculations \cite{effective, Sergei, ali, ali2}. Then we focus on how to include fields on the world sheet that transform under arbitrary representations of the unbroken group, and on spinor representations in particular (section \ref{addingf}). We then elucidate how the apparent simplicity of the Volkov-Akulov action, and the un-guage fixed Green-Schwarz action, are due to the specific spinor representations involved and a specific local field redefinition (section \ref{complete}).

There are several reasons that motivate constructing a world sheet action which has the full Lorentz group nonlinearly realized, with massless fermionic degrees of freedom which are not necessarily Goldstone fermions of broken supersymmetry. Physically this situation may arise because of the conjectured non-supersymmetric fermionic symmetry \cite{finitevolume}, which is proposed in the context of large-N circle compactified QCD. The spontaneous breaking of such symmetries as well as Lorentz invariance by confining strings which exist there, would result in non-SUSY Goldstone fermions. Another physical possibility is the situation in which fermions (usually chiral) are localized on the brane due to some specific target space dynamics. In such context these fermions are not even Goldstone degrees of freedom, and because of their chiral nature, nonlinear Lorentz invariance is realized in a very nontrivial way  (seciton \ref{complete}).

Regardless of their physical origin, there are formal motivations as well. It is well appreciated that quantum mechanically integrable systems are rare and constitute a very special subset of all physical theories. The criterion for integrability in two dimensions is equivalent to the requirement of factorizability of the full S-matrix in terms of two-to-two scattering \cite{zamo}. This leads to the algebraic condition given by the Yang-Baxter equation, and to the existence of an infinite set of conserved charges. Indeed, this turns out to be very restrictive, and any model that we find to obey these criteria warrants additional investigation. To this end, if we consider the world sheet affective field theory of a string in flat background, we obtain the Nambu-Goto action, which when examined in the static gauge exhibits very nontrivial interactions. This begs the question as to which underlying principle constrained the interaction vertices in such a way to ensure quantum integrability in $D=26$. The answer turns out to be that the hidden symmetries associated with the nonlinearly realized Poincare group fully determine the action and thus ensure tree level integrability, and imposing loop integrability implies the critical dimension. Conversely if we impose tree level integrability on a model with massless bosonic excitations with derivative interactions, we uniquely reproduce Nambu-Goto and obtain an action that realizes Poincare symmetry nonlinearly \cite{Sergei}. This discussion can be extended to the supersymmetric Poincare group, and one indeed finds the same phenomenon where space time hidden symmetries ensure integrability and that further restricts the target space dimensions to three, four, six and ten, already at tree level \cite{ali2}. In section \ref{critical} we will find out that if we start with a nonlinear sigma model of the Poincare group and add fermions as covariantly transforming matter fields, we have enough freedom to trivially ensure tree level integrability for fermions. However, the fermionic contribution to the one loop four boson scattering is fixed, which allows us to find model independent restrictions on the fermionic representations and target space time dimension. As a special case of this characterization, we find that one loop integrability singles out $D=10$ for supersymmetric models, with the fermionic representations corresponding exactly to the fermionic degrees of freedom in the static gauge of Green-Schwarz, Ramond-Neveau-Schwartz and Heterotic superstrings.

\section{Space-time CCWZ}
\subsection{The Algebra}\label{1}
For the $ISO(1,D-1)/ISO(1,1)\times SO(D-2)$ coset, we consider the Poincare algebra for the full group 

\begin{align}
[J^{\mu\nu},J^{\rho\sigma}] &= i \eta^{\mu\rho} J^{\nu\sigma} - i \eta^{\nu\rho} J^{\mu\sigma} + i \eta^{\nu\sigma} J^{\mu\rho} - i \eta^{\mu\sigma} J^{\nu\rho} \\
[J^{\mu\nu},P^{\rho}] &= i\eta^{\mu\rho}P^{\nu} - i\eta^{\nu\rho}P^{\mu}\\
[P^{\mu},P^{\nu}] &= 0
\end{align}
We have four types of generators:

\begin{table}[H]
\centering 
\begin{tabular}{c c c } 
\hline
Generator &    & Number  \\ [0.5ex] 
\hline 
$P^i$			&	broken translations			& $D-2$  					\\
$P^a$			&	unbroken translations 		& $2$ 					\\
$J^{ab} \,\&\, J^{ij}$	&	unbroken rotations/boosts 	& $1 + \frac{(D-2)(D-3)}{2}$ 	\\
$J^{ai}$			&	broken rotations/boosts		& $2(D-2)$				\\[1ex] 
\hline 
\end{tabular}
\end{table}
Here $a=\pm$ and $i,j={2,\ldots,D-2}$. 

Their algebra is similar to the Cartan Decomposition because schematically (here $X$ and $T$ stand for broken and unbroken generators respectively)
\begin{equation} 
[X,T] \propto X \qquad [T,T] \propto T \qquad [X,X] \propto X+T \label{cartandecomposition}
\end{equation}
Corresponding to broken translations we have $D-2$ physical Goldstone Bosons $X_{i}(\sigma)$, and corresponding to broken rotations we have $2(D-2)$ auxiliary Goldstone bosons $\phi_{ai}(\sigma)$. The reason why the latter are auxiliary rather than dynamical fields is because even though the broken generators are linearly independent, the low amplitude long wavelength excitations they generate need not be. In particular all the dynamical Goldstone modes correspond to physical excitations of the world sheet in the transverse directions. These are all accounted for by the broken translations, whereas a local broken rotation can always be represented as a combination of broken local translations \cite{manohar}.

A general element of the quotient group in the exponential parametrization \cite{salam} can be represented by
\begin{equation}\label{exponentialparametrization}
g(\sigma,X(\sigma),\phi(\sigma))= e^{i\sigma^{a}P_{a}+iX_{i}(\sigma)P_{i}}e^{i\phi_{ai}(\sigma)J_{ai}}
\end{equation}
Now since the variation of an element on the group manifold is the product of that element with an element of the tangent (Lie) space we can express the Cartan Form as
\begin{align}
g^{-1}dg \equiv ie_{a}&(X,\phi)P^{a}\,+\,iD_{i}(X,\phi)P^{i}\,+\,i\mathcal{V}^{ij}(X,\phi)J_{ij}^{(R2)}\nonumber\\&+\,i\mathcal{U}(X,\phi)J_{+-}^{(R1)}\,+\,i\Phi_{ai}(X,\phi)J_{ai}\label{cartanform}
\end{align}
Now using the defining eq(\ref{cartanform}) and eq(\ref{exponentialparametrization}) and the Poincare algebra with repeated application of the Jacobi identities one can find the dependence of $D,\mathcal{U},\mathcal{V},\Phi$ and $e$ on $X(\sigma)$ and $\phi(\sigma)$.

\subsection{Covariant Functions and Gauge Fields}
We are interested in the coefficient functions in the Cartan form eq(\ref{cartanform}) because they transform covariantly under the full group, and in particular under the broken generators. Under the unbroken generators. All Goldstones transform linearly under the unbroken generators. This can be immediately deduced from the fact that the Goldstone fields have the same quantum numbers as the broken generators, and the latter transform linearly under the unbroken generators as a consequence of eq(\ref{cartandecomposition}).

To deduce these transformations one notes the following: The action (left product) of any group element on an element of the quotient group, is another element of the quotient group. This new element does not have to be one of the representatives that we chose to parametrize the equivalence classes (in our case exponential parametrization). However since all elements inside an equivalence class are related through multiplication by the unbroken group, there must be an element of the latter that brings us back to the desired representative. Therefore
\begin{equation}
g_{1}g(\sigma,\xi(\sigma))=g(\widetilde{\sigma},\widetilde{\xi}(\widetilde{\sigma}))e^{iu(\sigma,\xi;g_{1})J_{+-}^{(R1)}+iv^{ij}(\sigma,\xi;g_{1})J_{ij}^{(R2)}} \label{transformation}
\end{equation}
Where
\begin{equation}
e^{iu(\sigma,\xi;g_{1})J_{+-}+iv^{ij}(\sigma,\xi;g_{1})J_{ij}}\equiv \Omega\subset SO(1,1)\times SO(D-2)
\end{equation}
is the restoring element, $\xi(\sigma)$ denotes both $X(\sigma)$ and $\phi(\sigma)$, and $g_{1}$ is a global Poincare transformation. The transformed coordinates and Goldstone bosons are referred to as $\widetilde{\sigma}$ and $\widetilde{\xi}(\widetilde{\sigma})$.\\
From eq(\ref{transformation}) and the transformation of the generators themselves under $\Omega$, we deduce the transformation of the Cartan form

\begin{align}
\widetilde{g}^{-1}d\widetilde{g} & =\Omega gdg\Omega^{-1}+\Omega d\Omega^{-1} \nonumber \\
	   & = i\left(\Lambda(u)^{a}_{\,\,a'}e^{a'}\right)P_{a} + i\left(R(v)^{j}_{j'}D^{j'}\right)P_{j} +i \left(\Lambda(u)^{a}_{\,\,a'}R(v)^{j}_{j'}\Phi^{a'j'}\right)J_{aj} \nonumber \\ &\qquad+ i\left(R(v)^{j}_{j'}R(-v)^{j}_{j'}(\mathcal{V}^{i'j'}-dv^{i'j'})\right)J_{ij}^{(R2)} + i \left(\mathcal{U}-du\right)J_{+-}^{(R1)}
\end{align}
where $\Lambda(u)^{a}_{\,\, a'}=\left(\text{exp}\left( iuJ_{+-} \right)\right)^{a}_{\,\,a'}$ is a boost in the fundamental representation of $SO(1,1)$, and $\mathcal{R}(v)^{i}_{j}=\left( \text{exp}\left( iv^{kk'}J_{kk'} \right)\right)^{i}_{j}$ is a rotation in the fundamental representation of $SO(D-2)$, and $u,v^{ij}$ are the functions defined by eq(\ref{transformation}). Thus expanding the one-forms we finally get
\begin{align}
&\widetilde{e}^{a}_{\alpha}(\widetilde{\sigma},\widetilde{\xi}(\widetilde{\sigma})) = \frac{\partial\sigma^{\alpha'}}{\partial\widetilde{\sigma}^{\alpha}}\Lambda(u(\sigma,\xi;g_{1}))^{a}_{\,\,a'}e^{a'}_{\alpha'}(\sigma,\xi(\sigma)) \nonumber \\
&\widetilde{D}^{j}_{\alpha}(\widetilde{\sigma},\widetilde{\xi}(\widetilde{\sigma})) = \frac{\partial\sigma^{\alpha'}}{\partial\widetilde{\sigma}^{\alpha}}R(v(\sigma,\xi;g_{1}))^{j}_{j'}D^{j'}_{\alpha'}(\sigma,\xi(\sigma)) \nonumber \\
&\widetilde{\mathcal{U}}_{\alpha}(\widetilde{\sigma},\widetilde{\xi}(\widetilde{\sigma})) = \frac{\partial\sigma^{\alpha'}}{\partial\widetilde{\sigma}^{\alpha}}\left(\mathcal{U}_{\alpha'}(\sigma,\xi(\sigma))-\partial_{\alpha'}u(\sigma,\xi;g_{1})\right) \nonumber \\
&\widetilde{\mathcal{V}}^{ij}_{\alpha}(\widetilde{\sigma},\widetilde{\xi}(\widetilde{\sigma})) = \frac{\partial\sigma^{\alpha'}}{\partial\widetilde{\sigma}^{\alpha}}R(v(\sigma,\xi;g_{1}))^{j}_{j'}R(-v(\sigma,\xi;g_{1}))^{j}_{j'}(\mathcal{V}(\sigma,\xi(\sigma))^{i'j'}_{\alpha'}-\partial_{\alpha'}v^{i'j'}(\sigma,\xi;g_{1})) \nonumber \\
&\widetilde{\Phi}^{aj}_{\alpha}(\widetilde{\sigma},\widetilde{\xi}(\widetilde{\sigma})) = \frac{\partial\sigma^{\alpha'}}{\partial\widetilde{\sigma}^{\alpha}}\Lambda(u(\sigma,\xi;g_{1}))^{a}_{\,\,a'}R(v(\sigma,\xi;g_{1}))^{j}_{j'}\Phi^{a'j'}_{\alpha'}(\sigma,\xi(\sigma)) \label{transformations2}
\end{align}
\paragraph{Frame field and Goldstone derivative\\}
By looking at eq(\ref{transformations2}) we observe that $D^{j}_{\alpha}$ provides us with covariantly transforming Goldstone boson field ``derivatives", and that $e^{a}_{\alpha}$ provides us with the frame field. 

Expanding the Cartan Form eq.(\ref{cartanform}) and using the Baker-Campbell-Haussdorff lemma (see appendix \ref{appendixA})  we deduce the frame field
\begin{equation}
e_{\alpha a} = \eta_{\alpha a} + \left( \phi^T \left( \frac{\text{cos }\sqrt{\phi\phi^T}-\mathds{1}}{\phi\phi^T}  \right)\phi \right)_{\alpha a} - \left( \phi \left( \frac{\text{sin }\sqrt{\phi\phi^T}}{\sqrt{\phi\phi^T}}  \right) \right)_{aj}\partial_{\alpha}X^{j} \label{framefield}
\end{equation}
And the Goldstone field covariant derivative 
\begin{equation}
D^{j}_{\alpha} = \partial_{\alpha}X_{i} \left(\text{cos }\sqrt{\phi\phi^T}\right)^{ij} + \left( \phi \left( \frac{\text{sin }\sqrt{\phi\phi^T}}{\sqrt{\phi\phi^T}}  \right)\right)^{j}_{\alpha}\label{implicit}
\end{equation}

We also need an invariant measure, which we readily obtain from the frame field
\begin{equation}
d^{2}\sigma\text{ det }e=d^{2}\sigma \sqrt{\text{det}\left(e^{a}_{\alpha}\,\eta_{ab}\,e^{b}_{\beta}\right)} 
\end{equation}

\paragraph{Spin-Connection and Gauge Field\\}
By examining the remaining equations in eq(\ref{transformations2}) we see that the non-homogeneous terms in the transformation of $\mathcal{U}_\alpha$ and $\mathcal{V}_\alpha^{ij}$ give us the exact transformations of a spin-connection and a gauge field respectively. Now we turn to eq(\ref{framefield}) and eq(\ref{implicit}) to write $\mathcal{U}$ and $\mathcal{V}$ in terms of the Goldstone fields. However this doesn't seem to produce a simple closed expression for the spin connection and the gauge field. We can nonetheless write down a recursive relation for the expansion in powers of $\varphi$, $\mathcal{U}_\alpha = \sum_n \mathcal{U}^{(n)}$ and same for $\mathcal{V}_{ij}$ where $\mathcal{U}^{(n)}$ and $\mathcal{V}^{(n)}$ are of order $\mathcal{O}(\varphi^n)$ .
\begin{align}
\mathcal{U}^{(n+2)}_\alpha &=\tfrac{i^n}{(n+4)(n+3)} \left(- \varphi^2 \,\mathcal{U}^{(n)}_\alpha + \epsilon^{ab}\varphi^j_a  \varphi ^k_b \,\mathcal{V}^{(n)}_{jk,\alpha} \right)
\label{UV1} \\
\mathcal{V}^{(n+2)}_{ij,\alpha} &= \tfrac{i^n}{(n+4)(n+3)} \left( \epsilon^{ab}\varphi_{aj}  \varphi_{bi} \,\mathcal{U}^{(n)}_\alpha + \varphi_i ^a \varphi^k_a \,\mathcal{V}^{(n)}_{jk,\alpha}\right)
\label{UV2}
\end{align}
with $\mathcal{U}^{(0)}_\alpha =- \frac{1}{2}\epsilon_{ab}\varphi^{aj}\partial_\alpha \varphi^{b}_j$, and $\mathcal{V}_{ij,\alpha}^{(0)} = -\frac{1}{2}\varphi^a_{[i}\partial_\alpha \varphi_{j]a}$.

\paragraph{Higher derivative terms\\}
To build higher derivative terms involving Goldstone fields only let us consider the following two derivatives
\begin{align}
(\nabla^{(1)}_{\alpha})^{b}_{a} &\equiv \delta^{b}_{a}\partial_{\alpha} - i(J_{+-})^{b}_{a}\mathcal{U}_{\alpha} \\
(\nabla^{(2)}_{\alpha})^{k}_{j} &\equiv \delta^{k}_{j}\partial_{\alpha} - i \mathcal{V}^{k}_{j\,\,\alpha}
\end{align}
These are special cases of the covariant matter field derivative operator (see section \ref{MF}) when the matter field is singlet under either unbroken subgroup.

We use these derivatives to construct the world sheet two form second rank $SO(1,1)$ tensor 
\begin{equation}
R_{\alpha\beta}^{ab} \equiv i[\nabla^{(1)}_{\alpha},\nabla^{(1)}_{\beta}]^{ab} = (J_{+-})^{ab}\partial_{[\alpha}\mathcal{U}_{\beta]}
\end{equation} 

and the world sheet two form second rank $SO(D-2)$ tensor 
\begin{equation}
(F_{\alpha\beta})^{k}_{j} \equiv i[\nabla^{(2)}_{\alpha},\nabla^{(2)}_{\beta}]^{k}_{j} = \partial_{[\alpha}\mathcal{V}^{k}_{\beta]j}-i[\mathcal{V}_{\alpha},\mathcal{V}_{\beta}]^{k}_{j} 
\end{equation}

We finally identify the world world sheet one-form $\Phi^{aj}_{\alpha}$ from eq(\ref{transformations2}) as a mixed $SO(1,1)$ and $SO(D-2)$ tensor.

These are useful for constructing higher order derivative interactions. The fact that these correspond to higher order derivatives becomes clear after solving the inverse Higgs constraint (section \ref{projecting}) where we will see that the auxiliary field is expressible in terms of the derivatives of the Goldstone fields. These higher order derivative terms correspond to higher geometric invariants (extrinsic curvature terms \cite{polyakov,kleinert}).\\

Notice that one can use the ``tetrad postulate" $\nabla^{(1)}_\alpha e^{\beta}_a = 0$ and the frame field eq(\ref{framefield}), to define a spin connection $\overline{\mathcal{U}}\left(\varphi,\partial X\right)$, namely
\begin{equation}\label{spinconnectionvielbein}
\begin{split}
\epsilon^{ab}\overline{\mathcal{U}}_\alpha =  &
\frac{1}{2} e^{a\beta}\left( \partial_\alpha e_{\,\,\,\beta}^b - \partial_\beta e_{\,\,\,\alpha}^b\right) 
- \frac{1}{2} e^{b\beta} \left( \partial_\alpha e_{\,\,\,\beta}^a - \partial_\beta e_{\,\,\,\alpha}^a\right) \\&\qquad\qquad 
- \frac{1}{2} e^{a\gamma}e^{b\delta} e_{\,\,\,\alpha} ^c\left(\partial_\gamma e_{\delta c} - \partial_\delta e_{\gamma c} \right)
\end{split}
\end{equation}
This is different from the one obtained by the standard CCWZ procedure eq(\ref{UV1}), however it transforms appropriately by construction. This is not surprising because the auxiliary field provides extra ingredients with which to build an object (the spin connection) that we postulate should transform according to eq(\ref{transformations2}).

Now that would give rise to a different world-sheet curvature tensor, which again transforms as a world-sheet two form second rank $SO(1,1)$ tensor
\begin{equation}
\overline{R}_{\alpha\beta}^{ab} \equiv  (J_{+-})^{ab}\partial_{[\alpha}\overline{\mathcal{U}}_{\beta]}
\end{equation} 

Nonetheless, when we project out the auxiliary fields, both $\mathcal{U}_\alpha\left(\varphi\right)$ and $\overline{\mathcal{U}}_\alpha\left(\varphi,\partial X\right)$ converge to the same object $\mathcal{U}_\alpha\left(\partial X\right)$. This cannot depend on our choice of constraint, as is to be expected because now we only have the right number of physical fields to construct an appropriately transforming spin connection. This is consistent with the fact that the geometric invariants $F$ and $R$ are unique, regardless of our derivation.

\subsection{Projecting out the Auxiliary Fields}\label{projecting}
The final step is to impose covariant conditions that result in finding the auxiliary fields in terms of the physical ones; $\phi^{aj}(\sigma)=\phi^{aj}(\sigma,X(\sigma))$.\\
The simplest $2(D-2)$ covariant constraints that we can impose are
\begin{equation} \label{constraintt}
D_{\alpha}^{j}=0
\end{equation}

So to obtain the correct relation between auxiliary fields and physcial ones we have to solve this set of nonlinear constraints. 

We notice that eq(\ref{transformations2}) are a set of dim$(ISO(1,D-1))$ equations, which in principle allows us to solve for $N(\bar{\xi})+$dim$(\mathbb{R}^{2})+N(u)+N(v^{ij})=$dim$(ISO(1,D-1))$ unknowns; that is to find $\widetilde{\xi}=\widetilde{\xi}(\sigma,\xi)$, $u=u(\sigma,\xi)$, $v=v(\sigma,\xi)$ and $\widetilde{\sigma}=\widetilde{\sigma}(\sigma,\xi)$. We will carry out this procedure in section \ref{tomf}.\\
Alternatively and more practically, we can use eq(\ref{transformation}) to deduce the same set of unknowns. 

\paragraph{Note:\\}
This covariant condition that projects out the auxiliary field coincides with the solution of the equation of motion of the latter in the case of det $e$ action (Nambu-Goto), however when we include matter fields this is not the case anymore. Then we can either stick to the simplest constraint eq(\ref{constraintt}), or alternatively use the equations of motion for the auxiliary field, which will depend on the particular interactions in the Lagrangian. In such case, the auxiliary field will be a function of both the Goldstone fields and the matter fields. Such a theory should eventually be equivalent to the one obtained by imposing the simple constraint eq(\ref{constraintt}) up to a field redefinition. For purposes of practicality it is clearly more convenient to choose the former constraint. However as we will see in (sec \ref{GSAction}), the latter will be more useful in the particular situation where we want to use $\kappa$-symmetry to re-derive the nonlinear Lorentz transformation.

If we use this relation in eq.(\ref{framefield}) we can write the frame field in terms of the physical bosons \footnote{If we use the constraint that follows from the equations of motion of the auxiliary fields instead, then eq(\ref{ff}) should be interpreted as the zeroth order in fermions} (see appendix \ref{appendixA}) 
\begin{equation}
e_{a\alpha} = \eta_{a\alpha} + \partial_{a}X^{i}\partial_{\alpha}X^{j} \left[  \left( (\partial X)^{T}\partial X \right)^{-1}\left( \sqrt{\mathds{1}+(\partial X)^{T}\partial X} -\mathds{1}\right)\right]_{ij} \label{ff}
\end{equation}
Where $(\partial X)^{T}\partial X$ is the $(D-2)\times(D-2)$ matrix $\partial_{a}X^{i}\partial^{a}X^{j}$. This gives the expected induced metric 
\begin{equation}
h_{\alpha\beta} = e_{a\alpha}\eta^{ab}e_{b\beta} = \eta_{\alpha\beta} + \partial_{\alpha}\vec{X} \cdot \partial_{\beta}\vec{X}
\end{equation}

\section{Adding Fermions to the World Sheet}\label{addingf}
\subsection{Matter Fields}\label{MF}
Now we introduce a covariantly transforming matter field in the $R1\otimes R2$ representation
\begin{equation}
\widetilde{\psi}_{aj}(\widetilde{\sigma}) = \mathcal{D}^{(R1)}(\Lambda(u))^{a'}_{a}\mathcal{D}^{(R2)}(\mathcal{R}(v))^{j'}_{j}\psi_{a'j'}(\sigma) \label{matterfield}
\end{equation}
where $\mathcal{D}^{(R1)}\left(\Lambda(u)\right)^{a}_{\,\, a'}=\left(\text{exp}\left( iuJ_{+-}^{(R1)} \right)\right)^{a}_{\,\,a'}$ is a boost in the $R1-$representation of $SO(1,1)$, and $\mathcal{D}^{(R2)}\left(\mathcal{R}(v)\right)^{i}_{j}=\left( \text{exp}\left( iv^{kk'}J_{kk'}^{(R2)} \right)\right)^{i}_{j}$ is a rotation in the $R2-$representation of $SO(D-2)$.
We have a covariantly transforming matter field derivative 
\begin{equation}
\left(\widetilde{\nabla}_{\alpha}\widetilde{\psi}(\widetilde{\sigma})\right)_{aj} = \frac{\partial \sigma^{\alpha'}}{\partial \widetilde{\sigma}^{\alpha}}\mathcal{D}^{(R1)}(\Lambda(u))^{a'}_{a}\mathcal{D}^{(R2)}(\mathcal{R}(v))^{j'}_{j}\left(\nabla_{\alpha'}\psi(\sigma)\right)_{a'j'}\nonumber
\end{equation}
when we define
\begin{equation}
\left(\nabla_{\alpha}\right)^{k\,\,b}_{j\,\,a} = \delta^{k}_{j}\delta^{b}_{a}\partial_{\alpha}-i\delta^{k}_{j}\left(J_{+-}^{(R1)}\right)^{b}_{a}\mathcal{U}_{\alpha}-i\delta^{a}_{b}\mathcal{V}^{ll'}_{\alpha}\left( J^{(R2)}_{ll'} \right)^{k}_{j} \label{covder}
\end{equation}
Notice that because $SO(1,1)$ is an Abelian subgroup, $J_{+-}$ will be diagonal (e.g. diag$\{\frac{1}{2},-\frac{1}{2}\}$ for Dirac), and will be just a c-number for irreducible representations.

For the world sheet Dirac spinor we can now introduce invariant terms in the Lagrangian, e.g.
\begin{equation}
i\bar{\psi}\rho^{a}e^{\alpha}_{a}\nabla_{\alpha}\psi 
\qquad 
i\bar{\psi}\rho^{*}\rho^{a}e^{\alpha}_a\nabla_{\alpha}\psi \label{inv} 
\qquad 
i\bar{\psi} e^\alpha_a\rho^a \rho_b  \psi \Phi^b_\alpha \qquad \ldots
\end{equation}
Where  $\rho$ are gamma matrices satisfying Clifford algebra in flat two dimensional space $\{ \rho^a, \rho^b\} = 2\eta^{ab}$, and $\rho^*$ is the $2D$ chirality operator $\rho^0\rho^1$.
\paragraph{Note:} while a Lagrangian of the form $\mathcal{L} = i \,\text{det }e\left(1+ e^\alpha_a \bar{\psi} \rho^a \nabla_\alpha\psi\right) $ is not hermitian, one can use the tetrad postulate after projecting out the auxiliary fields\footnote{Before imposing the inverse Higgs constraint, the tetrad postulate holds for $\overline{\mathcal{U}}$ and not for $\mathcal{U}$. But as discussed in the previous section this distinction disappears after imposing the constraint.} to show that up to a total derivative it is equivalent to the hermitian Lagrangian
\begin{equation}
\mathcal{L} = i \,\text{det }e +i\,\text{det }e \, e^\alpha_a \bar{\psi} \rho^a \left( \overset{\leftrightarrow}{\partial}_\alpha + i J_{ij}^{(R2)} \mathcal{V}^{ij}_\alpha \right)\psi
\label{UVV}
\end{equation}
This Lagrangian is hermitian, and has no reference to the spin connection at all. This is a special feature of the two dimensional case where the spin connection term vanishes because $\rho^*$ is the generator of world-sheet rotations and $\{ \rho^*, \rho^a\}=0$.

\subsection{Transformation of Matter Fields}\label{tomf}
Now we take a closer look at the transformation of matter fields given in eq(\ref{matterfield}). We observe that the transformation of a matter field under an element $g_1$ of the full group is determined by the functions $u(\varphi(\sigma), g_1)$ and $v^{ij}(\varphi(\sigma), g_1)$ appearing in eq(\ref{transformation}). We already know that when $g_1\in SO(1,1)\times SO(D-2)$, $u$ and $v^{ij}$ are just constants because the fields transform linearly. However for broken generators we need to find $u_{ak} \equiv u(\varphi(\sigma),J_{ak})$ and $v^{ij}_{ak}\equiv v^{ij}(\varphi(\sigma),J_{ak})$ as they will now be functions of the auxiliary fields. To do so we use eq(\ref{transformation}) to find

\begin{align}
 & e^{-i\phi_{\beta i}J^{\beta i}} \left( P_{\alpha '} \left( \delta^{\alpha i}_{\theta}\sigma^{\alpha'} + \theta\eta^{\alpha\alpha'}X^{i} \right) + P_{j'} \left( \delta^{\alpha j}_{\theta}X^{j'} - \theta\sigma^{\alpha}\eta^{jj'} \right) \right) e^{+i\phi_{\beta i}J^{\beta i}} = \nonumber \\
 & \qquad\qquad e^{-i\phi_{\beta i}J^{\beta i}} \left( \theta J^{\alpha j} - \delta^{\alpha j}_{\theta} \right) e^{+i\phi_{\beta i}J^{\beta i}}  - \theta u^{\alpha j}B -\theta v^{\alpha j}_{ii'}J^{ii'}
\end{align}

Since the right hand side only has momentum generators and the left hand side only rotations, both must be zero. Which implies the nonlinear transformation for coordinates and bosons
\begin{equation}
\delta^{\alpha i}_{\theta}\sigma^{\alpha'} = - \theta\eta^{\alpha\alpha'}X^{i} \qquad
\delta^{\alpha j}_{\theta} X^{j'} = \theta \sigma^{\alpha}\eta^{jj'} \label{wellknown}
\end{equation}
This transformation law could have been deduced from upgrading the world sheet coordinates to fields to introduce reparametrization invariance \cite{cooper}, then we can arrange all the bosonic fields to be linearly transforming under the vector representation of $SO(1,D-1)$, in that picture one verifies eq(\ref{wellknown}) as the compensating diffeomorphism that ensures the static gauge. That is because broken boosts/rotations do not preserve the static gauge, unlike the unbroken transformations which do.

The right hand side gives a coupled set of equations, the coefficient of $B$ and $J^{ij}$ gives us $u^{\alpha j}(\varphi)$ and $v^{\alpha j}_{ii'}(\varphi)$ in terms of $\delta^{aj}\varphi$ respectively, and the coefficient of $J^{aj}$ allows us to solve for $\delta^{aj}\varphi$. These again do not seem to easily lend themselves to a closed expression (see appendix \ref{appendixA}) , but $u$ and $v$ can be written recursively in very similar way to eq(\ref{UV1}) and eq(\ref{UV2}).

We'll write the leading order solution; from $J^{aj}$ we get  $\delta^{aj}\varphi^{b}_k = \eta^{ab}\delta^j_k +\mathcal{O}\left(\varphi\right)$, which implies 
\begin{equation}\begin{split}
u^{aj}(\varphi) & = \epsilon^{ba}\varphi_b^j - \frac{1}{2} \varphi^i_b \delta^{aj}\varphi_{ci} \epsilon^{bc} + \ldots = \frac{1}{2}  \epsilon^{ba}\varphi_b^j  + \mathcal{O}(\varphi^2) \\
v^{aj}_{ik}(\varphi) &= \varphi ^a_{[i}\delta^j_{k]} - \frac{1}{2} \varphi^b_{[i} \delta^{aj}\varphi_{k]b} + \ldots = \frac{1}{2}\varphi ^a_{[i}\delta^j_{k]} + \mathcal{O}(\varphi^2)
\end{split}\end{equation}

In section \ref{projecting} we saw how we can write the auxiliary field $\varphi$ as an expansion in both the Goldstone fields as well as the matter fields. To leading order however we can write $\varphi^i_a = \partial_a X^i + \mathcal{O}\left( \left(\partial X\right)^2, \bar{\psi}\psi\right)$ so that 
\begin{equation}\begin{split}
u^{aj} & = \frac{1}{2}  \epsilon^{ba}\partial_b X^j  + \ldots \\
v^{aj}_{ik} &=  \frac{1}{2}\partial ^aX _{[i}\delta^j_{k]} + \ldots
\end{split}\end{equation}

There for we can now write down the infinitesimal form of the transformation under the broken generators in eq(\ref{matterfield}) to leading order in derivatives and fermions as 
\begin{equation}
\delta^{aj}\psi  = \left( \frac{1}{2}\epsilon^{ba}\partial_b X_j \rho^*  + \partial^a X_k J^{ij}_{R} + \ldots \right)\psi \label{nonlineartransformationlaw}
\end{equation}

\section{Complete Spinors and the Volkov-Akulov Action}\label{complete}
Now we can compare the CCWZ construction of broken Poincare with fermionic matter fields to known supersymmetric theories that realize Poincare nonlinearly. It goes without saying that the latter should be a subcategory of the former. However to make this connection apparent, we must deal with the fact that the fermionic fields in supersymmetric theories such as the Volkov-Akulov action for completely broken supersymmetry, appear to transform linearly under the full Poincare group, whereas we would expect them to transform linearly only under the unbroken subgroup according to the canonical CCWZ procedure.

To reconcile these two pictures, one can consider the more general possibility of adding enough matter fields that can be arranged into a multiplet of the full group. According to the CCWZ construction, each of the irreducible components of this multiplet transforms nonlinearly under the broken generators and linearly under the unbroken. This set of theories is larger than those with matter fields that form multiplets under the full group as can be seen for instance by the additional free parameters that appear in the Lagrangian.  Now if the issue of uniqueness of the CCWZ construction for space time symmetries is not relevant in this context, it follows that the latter set of theories should be physically equivalent to a subset of the former. The reason being that although these two multiplets transform differently under the broken generators, they both are identical from the point of view of the unbroken subgroup (low-energy effective theory), and commutation relations of the Poincare transformations satisfy the same $SO(1,D-1)$ algebra.

It follows then that in our case there must exist a field redefinition which relates the two fields.

To be concrete let's consider a Weyl spinor $\Psi$ in some even dimension $D$. Then we know that for the unbroken generators this representation splits into the sum of two irreducible ones as 
\begin{equation}
\textbf{2}^{D/2 - 1}  \simeq \left( \textbf{1}_{+}\otimes \textbf{2}^{D/2-2}_{+}\right)\oplus \left( \textbf{1}_{-}\otimes\textbf{2}^{D/2-2}_{-}\right) \label{Clebsch}
\end{equation}
So according to the discussion above, let us consider left and right handed Weyl spinors $\psi_{\pm j}$ on the world sheet transforming under $J^{aj}$ as
\begin{align}
	\tilde{\psi}_{+ k} &=  e^{+ \theta u^{aj}(\partial X)} \text{exp} \left( -\tfrac{\theta}{8} \left(\sigma^{i}\bar{\sigma}^{i'}-\sigma^{i}\bar{\sigma}^{i'}\right)v_{ii'}^{aj}(\partial X)   \right)_{k}^{\,\,k'} \psi_{+k'} \\
	\tilde{\psi}_{- k} &=  e^{- \theta u^{aj}(\partial X)} \text{exp}\left( +\tfrac{\theta}{8} \left(\bar{\sigma}^{i}\sigma^{i'}-\bar{\sigma}^{i}\sigma^{i'}\right)v_{ii'}^{aj}(\partial X)   \right)_{k}^{\,\,k'} \psi_{-k'}	
\end{align}
where $\sigma^{i}$ are the $D-2$ dimensional Pauli matrices. Then the objective is to find the field redefinition that mixes the components of these two spinors into a new spinor which transforms linearly under both broken and unbroken transformations.

The desired field redefinition is given by
\begin{equation}\label{redefinition}
	\tilde{\psi}^{\alpha k} = \left(  \text{exp}\left(  \tfrac{1}{4}\phi_{aj}\rho^*\rho^a\otimes\gamma^j    \right)\right)^{\alpha\beta,kk'} \psi_{\beta k'}
\end{equation}
where $\phi_{aj}$ as defined in eq(\ref{exponentialparametrization}), and the D-dimensional Clifford algebra is used in the following form
\begin{align}
	\Gamma^{a} &= \rho^{a}\otimes\mathds{1}_{D/2-1} \\
	\Gamma^{i}  &= \rho^{*}\otimes \gamma^{i}
\end{align}	
where $\gamma^i$ are gamma matrices in $D-2$ dimensions, and $\rho^a$ are the two-dimensional gamma matrices which we choose to be in the real Weyl representation 
\begin{align}
\rho^0 = \left( \begin{array}{cc} 0 & -1 \\ 1 & 0 \end{array}\right) \qquad \rho^1 = \left( \begin{array}{cc} 0 & 1 \\ 1 & 0\end{array} \right) \qquad \rho^* = -\rho^0 \rho^1 = \left( \begin{array}{cc} 1 & 0 \\ 0 & -1 \end{array}\right) 
\end{align}
and $\rho^{\pm}\equiv \frac{1}{\sqrt{2}} \left( \rho^0 \pm \rho^1\right)$.
\\

In the original representation $\Psi^{A} \equiv C^{A}_{\alpha k}\tilde{\psi}^{\alpha k}$, where $C^{A}_{\alpha k}$ are the Clebsch-Gordon coefficients that establish the correspondence in eq(\ref{Clebsch}). Or equivalently, in the original bases we write
\begin{equation}
	\tilde{\Psi} = e^{  \phi_{aj}\Gamma^a \Gamma^j /4   }\Psi
\end{equation}

Notice that this field redefinition is nonlinear, that is because as discussed in section \ref{projecting}, $\phi$ itself is a function of both $\partial X$ and $\psi$.

As an example of these complete spinors, we consider a string in supersymmetric Minkowski background, and we also consider the case where the string breaks all supercharges. As far as nonlinearly realized Poincare symmetry is concerned, we may think of these Goldstinos as spinor matter fields. Furthermore, since all supercharges are broken, and they form a representation under the full Lorentz group, we know that the Goldstinos fall into a representation of the full Lorentz group, and that it is equivalent to say that they transform linearly or nonlinearly under the broken generators, up to a field redefinition.

Then it is instructive to see how the Volkov-Akulov action is an example of a nonlinearly realized Poincare with spinor matter fields.
\begin{equation}
S_{AV} = \int d^2\sigma \sqrt{\text{det }\Pi^{\mu}_{\alpha}\Pi_{\mu\beta}}
\end{equation}
Where $\Pi^{\mu}_{\alpha} \equiv \partial_{\alpha}X^{\mu} - i\bar{\psi}\Gamma^{\mu}\partial_{\alpha}\psi + h.c.$, and the action is written in a diffeomorphism invariant way which in according to our consideration should be fixed into the static gauge $X^{\mu}=(\sigma^{\alpha},X^{j})$. Then upon expanding the Lagrangian in powers of fermions and specializing to $D=3$, we obtain (see appendix \ref{appendixB}) 
\begin{align}
\mathcal{L}_{AV} 	&=  \mathcal{L}_{NG} \left(    2i\bar{\psi} \rho_\gamma \partial_\beta \psi \left(  \eta^{\beta\gamma}  - \frac{\partial^\beta X\partial^\gamma X}{1+\left(\partial X\right)^2}   \right)+ \frac{2i\bar{\psi} \rho^* \partial_\beta \psi}{1+\left(\partial X\right)^2} \partial^\beta X + h.c.  \right) +\mathcal{O}\left( (\bar{\psi}\rho\partial\psi)^2\right) \nonumber\\
&= \mathcal{L}_{NG} \left(    2i g^{\alpha\beta}\left(\partial X\right)\bar{\psi} \rho_\alpha \partial_\beta \psi + 2ig^{\alpha\beta}\left(\partial X\right)\bar{\psi} \rho^* \partial_\alpha \psi \partial_\beta X + h.c.  \right) +\mathcal{O}\left( (\bar{\psi}\rho\partial\psi)^2\right) \label{volkovd3}
\end{align}
Let's compare this to $SO(1,2)/SO(1,1)$ with a Dirac matter field. In appendix \ref{appendixC} we find the action to be eq(\ref{simplestaction})
\begin{align}	
 \mathcal{L} =\text{det }e \bigg[
1- \left(	\eta^{ab} - \varphi^a \varphi^b \tfrac{\text{cos}|\varphi| - 1}{\varphi^2} \right) e^\alpha_a \bar{\psi}\rho_b \partial_\alpha\psi  - \varphi^a \text{sinc}|\varphi| e^\alpha_a\bar{\psi}\rho^*\partial_\alpha\psi \bigg] 
\end{align}

Now we must project out the auxiliary fields, and we choose the simple Nambu-Goto inverse Higgs constraint $D_\alpha=0$ to obtain 
\begin{equation}
\varphi_\alpha = \partial_\alpha X \frac{\text{arctan}\sqrt{\left(\partial X\right)^2}}{\sqrt{\left(\partial X\right)^2}} 
\end{equation}
Substituting this in our CCWZ lagrangian indeed reproduces eq(\ref{volkovd3}).

Here, unlike if we were only concerned with nonlinearly realized Lorentz, the coefficient $\alpha_2' = -1$ in the lagrangian eq(\ref{simplestaction}) was fixed by nonlinearly realized supersymmetry.

\section{Equivalent Constructions}
Here we provide alternative derivations of the transformation rules for covariant matter fields under broken Lorentz transformations. First by using a very pragmatic approach based on the consistency of the commutation relation between the generators of the Lorentz algebra. This will provide a quick cross check for eq(\ref{nonlineartransformationlaw}). Second we adopt a far less practical approach, which relies on well established models with nonlinearly realizes supersymmetry. However what we loose with practicality we gain in insight, and we find how is the $\kappa$-symmetry of the Green-Schwarz action related to nonlinear Lorentz transformations. 
\subsection{Commutation Relations}

Were we only concerned with the first few terms in the small field expansion, we could have deduce the transformation of the matter fields under the broken generators eq(\ref{matterfield}) perturbatively, by imposing the commutation relations of the Poincare algebra at each order. And then deduce the form of the spin-connection accordingly.

To demonstrate the procedure let's consider the simple case of a Weyl spinor on the world sheet of a string propagating in 3 dimensional flat space time, that is the theory $SO(1,2)/SO(1,1)$.

Under the unbroken generator $J^{+-}$ the spinor transforms linearly as
\begin{equation}
\tilde{\psi}(\tilde{\sigma}) =  e^{+\theta/2}\psi(\sigma) 
\end{equation}
which infinitesimally gives
\begin{equation}
\delta^{+-}_{\theta} \psi(\sigma) = \tfrac{\theta}{2} \psi(\sigma) - \theta \epsilon^{\alpha\beta}\sigma_{\beta}\partial_{\alpha} \psi(\sigma) + \mathcal{O}(\theta^{2})
\end{equation}
Now we propose the most general form of the transformation under $J^{\alpha 2}$
\begin{equation}
\tilde{\psi}(\tilde{\sigma}) = \text{exp}\left\{ \theta    \left(\epsilon^{\alpha\beta}\partial_{\beta}X + c \partial^{\alpha}X \right)  \mathcal{F}\left(\left(\partial X\right)^{2}\right)     \right\} \psi(\sigma)
\end{equation}
Where we can expand $\mathcal{F}\left(\left(\partial X\right)^2 \right) = \mathcal{F}_o + \mathcal{F}_1 \left(\partial X\right)^2+\ldots$. 

Using $\delta^{\alpha 2}_{\theta} \sigma^{\beta} = -\theta \eta^{\alpha\beta} X$, this gives the infinitesimal transformation
\begin{align}
\delta^{\alpha 2}_{\theta} \psi(\sigma)  &=  \theta\mathcal{F}((\partial X)^{2}) \left(\epsilon^{\alpha\beta}\partial_{\beta}X + c \partial^{\alpha}X \right) \psi(\sigma) + \theta X \partial^{\alpha} \psi(\sigma) +\mathcal{O}(\theta^{2}) \nonumber\\
&= \theta \left(  \mathcal{F}_{0} + (\partial X)^{2} \mathcal{F}_{1} \right) \left(\epsilon^{\alpha\beta}\partial_{\beta}X + c \partial^{\alpha}X \right) \psi(\sigma) + \theta X \partial^{\alpha} \psi(\sigma)  + \mathcal{O}((\partial X)^{5}) \label{spinortrans}
\end{align}
Now we want to use the commutation relation $[J^{\alpha 2},J^{\beta 2}]=i\epsilon^{\alpha\beta}B$ to deduce the constants $\mathcal{F}_{0}$, $\mathcal{F}_{1}$ and $c$. In other words
\begin{equation}
[\delta^{+2}_{\theta},\delta^{-2}_{\theta}]\psi(\sigma) = \tfrac{\theta}{2} \psi(\sigma) - \theta \epsilon^{\alpha\beta}\sigma_{\beta}\partial_{\alpha} \psi(\sigma)
\end{equation}
Which gives us $\mathcal{F}_{0} = \frac{1}{2}$, $\mathcal{F}_{1} = -\frac{1}{8}$ and $c=0$, which is consistent with eq.(\ref{nonlineartransformationlaw}) for the dimension and representation at hand. Then it is straightforward to deduce the spin-connection once we have the transformation law.

\subsection{Green-Schwarz Action}\label{GSAction}
In the same way that we can deduce the transformation laws under the broken Lorentz generators of Goldstone fields eq(\ref{wellknown}) as a combination of a global Lorentz transformation with a compensating diffeomorphism, we can use the Green Schwarz action with $N=1$
\begin{equation}
S =\frac{1}{2\pi} \int d\sigma^{2} \sqrt{-h}\,h^{\alpha\beta}\Pi^{\mu}_{\alpha}\Pi_{\mu\beta}  - \frac{i}{\pi} \int d\sigma^{2} \epsilon^{\alpha\beta}\partial_{\alpha}X^{\mu}\left( \bar{\theta}\Gamma_{\mu}\partial_{\beta}\theta \right)  \label{gs}
\end{equation}
 (With $\Pi^{\mu}_{\alpha} \equiv \partial_{\alpha}X^{\mu} - i\bar{\theta}\Gamma^{\mu}\partial_{\alpha}\theta$ for Majorana) in dimension where it is defined to deduce the transformation laws of spinors in the corresponding specific representation. This furnishes the transformations of Majorana spinors for $D=3$, Majorana/Weyl for $D=4$, Weyl for $D=6$ and Majorana-Weyl for $D=10$, which in the physical gauge eq(\ref{gauge}) reduce to fields in the spinor representation $(\mathbf{1}_{-},\mathbf{2}^{D/2-2}_{-})$ of $SO(1,1)\times SO(D-2)$, see eq(\ref{Clebsch}). We will see that this way we will obtain a nonlinear realization for the matter fields, which will turn out to be equivalent to the broken Lorentz CCWZ for certain coefficients imposed by SUSY.

We proceed by noticing that both Goldstone bosons and the fermions transform linearly under the global Lorentz symmetry before fixing diffeomorphism invariance and $\kappa$-symmetry \cite{kappa}. However once we fix the physical gauge 
\begin{equation}
X^{\alpha}(\sigma) = \sigma^{\alpha}, \qquad\qquad  \Gamma^+ \theta \equiv \tfrac{1}{\sqrt{2}}\left( \Gamma^{0}+\Gamma^{D-1}\right)\theta= 0  \label{gauge}
\end{equation}
a general unbroken global Lorentz transformation violates it\footnote{Exponents in the unbroken global Lorentz transformations exp${\{+\omega_{\alpha\beta}[\Gamma^{\alpha},\Gamma^{\beta}]}/8\}$ and exp$\{{+\omega_{ij}[\Gamma^{i},\Gamma^{j}]}/8\}$ anti commute and commute with $\Gamma^{+}$ respectively, and thus the gauge fixing condition is preserved, whereas the exponent in exp$\{{+\omega_{\alpha i}[\Gamma^{\alpha},\Gamma^{i}]}/8\}$ doesn't.}, which means that we have to supplement global Lorentz transformations with compensating diffeomorphism and $\kappa$-symmetry transformations to ensure we remain in that gauge.

To demonstrate this procedure let's consider the simplest Green Schwarz action; $D=3$ with $N=1$ Weyl spinor. The diffeomorphism is
\begin{align}
\delta_{\xi} \sigma^{\alpha} &=   \xi^{\alpha} (\sigma) \label{one}  \\
\delta_{\xi} h^{\alpha\beta} &=  \partial^{(\alpha}\xi^{\beta)}(\sigma) \label{two}
\end{align}
which implies $ \delta_\xi X^\mu = \xi^\alpha \partial_\alpha X^\mu,\, \delta_\xi \theta = \xi^\alpha \partial_\alpha \theta$. 

The $\kappa$-symmetry transformation is\footnote{The transformation of $h^{\alpha\beta}$ in the static gauge doesn't provide any independent information, and must be consistent with the transformation of the fields and how the auxiliary metric is given in terms of them.}
\begin{align}
\delta_{\kappa} \theta &=  \Gamma_{\mu}\Pi^{\mu}_{\alpha} 
					\kappa^\alpha\label{three} \\
\delta_{\kappa} X^{\mu} &= i\bar{\theta} \Gamma^{\mu} \delta_\kappa\theta\label{four} 
\end{align}
with the duality condition\footnote{Even though this condition seems to imply two constraints on $\kappa$, in fact it is degenerate so we can consider either components of the condition, or a linear combination thereof.} (for $N=1$) 
\begin{equation}
\left( \sqrt{-h}h^{\alpha\beta} + \epsilon^{\alpha\beta}\right)\kappa_{\beta}=0 \label{condition}
\end{equation}

The gauge conditions eq(\ref{gauge}) in our case are
\begin{equation}
X^{\mu} = (\sigma^{\alpha},X),\qquad\qquad 
\theta = \left( \begin{array}{c}\theta_L \\ 0\end{array}\right)
\end{equation}
Under a broken Lorentz generator $J^{\alpha i}$ this gauge condition is violated as
\begin{align}
X^{\mu} &= (\sigma^{\beta} + \epsilon\eta^{\beta\alpha} X, X - \epsilon \sigma^{\alpha}) \nonumber \\
\Gamma^+ \theta &= -\tfrac{1}{4} \epsilon\Gamma^+ \Gamma^\alpha  \theta
\end{align}

This implies that under $J^{-2}$ the spinor is invariant, so no $\kappa$ compensation is required, whereas under $J^{+2}$
\begin{equation}
\tilde{\theta} = \left( \begin{array}{c} \theta_L \\ -\tfrac{\epsilon}{4}\theta_L \end{array}\right)\label{leftright}
\end{equation}

We should use the gauge transformations eq(\ref{one},\ref{three},\ref{four}) to restore eq(\ref{gauge}), which gives the conditions
\begin{align}
\left( \delta_\xi + \delta_\kappa \right) \sigma^\beta  &= - \epsilon \eta^{\beta\alpha}X \nonumber \\
 \Gamma^+ \left( \delta_\xi + \delta_\kappa \right) \theta &=  -\tfrac{\epsilon}{4} \Gamma^+ \Gamma^\alpha  \theta 
\end{align}

\paragraph{Solving for $\kappa$\\}
The duality condition eq(\ref{condition}) relates $\kappa_+$ and $\kappa_-$ by either of these two degenerate equations \footnote{The coefficients in these equations are not spinor matrices, but c-numbers}
\begin{align}
\Pi_- \cdot \Pi_- \kappa_+ - \left(\Pi_+ \cdot \Pi_- - \mathcal{L}_{V} \right)\kappa_- &= 0 \\
\Pi_+ \cdot \Pi_+ \kappa_- - \left(\Pi_+ \cdot \Pi_- + \mathcal{L}_{V} \right)\kappa_+ &= 0 
\end{align}
for both left and right components of the spinors $\kappa_\pm$. Where we used the fact that now $h^{\alpha\beta}$ is no more an independent field, but given by $h_{\alpha\beta}/\sqrt{-h} = \Pi_{\alpha}\cdot\Pi_{\beta}/\sqrt{-\text{det }\Pi_{\alpha}\cdot\Pi_{\beta}}\equiv\Pi_{\alpha}\cdot\Pi_{\beta}/\mathcal{L}_{V}$.
We use the first equation to write $\kappa_-$ in terms of $\kappa_+$
\begin{equation}
\kappa_- = \frac{\Pi_- \cdot \Pi_- }{\Pi_+ \cdot \Pi_- - \mathcal{L}_{V}}\kappa_+
\end{equation}
Which gives us now
\begin{align}
\delta_{\kappa} \theta &=  \hat{\Pi}_{\alpha}h^{\alpha +} \kappa_+ + \hat{\Pi}_{\alpha}h^{\alpha -} \kappa_- \\
&= \left(\hat{\Pi}_{\alpha}h^{\alpha +}\left(\Pi_+ \cdot \Pi_- - \mathcal{L}_{V}\right) + \hat{\Pi}_{\alpha}h^{\alpha -}\Pi_-\cdot \Pi_- \right) \frac{\kappa_+}{\Pi_+ \cdot \Pi_- - \mathcal{L}_{V}} \\
&= \left( a\,\hat{\Pi}_+ + b\,\hat{\Pi}_-\right)\frac{\kappa_+}{\Pi_+\cdot\Pi_- - \mathcal{L}_V}
\end{align}
Where $a$ and $b$ are the field dependent c-numbers
\begin{align}
a &=	-\Pi_-\cdot\Pi_- 			\\
b &=	 \Pi_+\cdot\Pi_- - \mathcal{L}_V 
\end{align}
One can verify that, det$\left( a\,\hat{\Pi}_+ + b\,\hat{\Pi}_-\right)=0$, so that this spinor matrix is singular. From which we immediately deduce that the $\kappa$ transformation $\delta_{\kappa}\theta_R = +\tfrac{\epsilon}{4}\theta_L$ acting on $\theta_R = -\tfrac{\epsilon}{4}\theta_L$ which restores the light cone gauge, also shifts $\theta_L$ by
\begin{equation}
\delta_{\kappa}\theta_L = +\tfrac{\epsilon}{4} \theta_L\frac{\left( a\,\hat{\Pi}_+ + b\,\hat{\Pi}_-\right)_{11}}{\left( a\,\hat{\Pi}_+ + b\,\hat{\Pi}_-\right)_{21}} = +\tfrac{\epsilon}{4} \theta_L\frac{\left( a\,\hat{\Pi}_+ + b\,\hat{\Pi}_-\right)_{12}}{\left( a\,\hat{\Pi}_+ + b\,\hat{\Pi}_-\right)_{22}}\equiv \tfrac{\epsilon}{4}\mathcal{K}\theta_L
\end{equation}
We only need to consider $\mathcal{K}$ to zeroth order in $\epsilon$, so all $\theta_R$ are ignored, then we use the fact that $\theta_{L}^2=0$ to retain only the $\partial X$ dependence
\begin{align}
\mathcal{K} &= \frac{		-\Pi_-\cdot\Pi-(1+i\theta_L\partial+\theta_L	) + i\theta_L\partial_L\theta_L(\Pi_-\cdot\Pi_+ - \mathcal{L}_V)}{-\Pi_-\cdot\Pi_- (-\partial_+X +  \mathcal{O}(\epsilon)) + (\Pi_-\cdot\Pi_+ - \mathcal{L}_V)(-\partial_- X + \mathcal{O}(\epsilon))}\nonumber \\
&\longrightarrow -\frac{\partial_- X}{1+\sqrt{1 + (\partial X)^2}}
\end{align}
Where $\mathcal{L}_V \longrightarrow \sqrt{1+(\partial X)^2}$ when we drop $\theta$ dependence.
Now we finally can write the transformation rules under all Lorentz generators
\begin{align}
\delta^{+-}_{\epsilon}\theta_L &=\tfrac{\epsilon}{2}\theta_L \\
\delta^{-2}_{\epsilon}\theta_L &=0\\
\delta^{+2}_{\epsilon}\theta_L &=-\frac{\epsilon}{4}\frac{\partial_- X}{1+\sqrt{1 + (\partial X)^2}}
\end{align}
Even though this nonlinear realization seems to be different from the one in eq(\ref{ud3}), one can show that a field redefinition as in eq(\ref{redefinition}) establishes their equivalence.  \\ 

Also notice that for less trivial examples where higher orders in fermions do not vanish, the compensating $\kappa$ transformation, owing to its nonlinearity, will itself be an expansion in terms of fermions. The analogue of this property in the CCWZ formalism comes from the fact discussed in section \ref{projecting}, namely that the auxiliary fields themselves will be nonlinear functions of the fermions when we choose to solve the field equations of the auxiliary field rather than imposing the simplest inverse Higgs constraint. So that when we perform a field redefinition as in eq(\ref{redefinition}), and obtain transformation laws in terms of $u^{\alpha j}(\varphi)$, all of this $\varphi$ dependence will embody the nonlinearity inherent to $\kappa$ transformations.

\section{The critical dimension $D=10$}\label{critical}

Now that we have the ingredients to construct a nonlinear sigma model of broken Poincare symmetry due to the presence of a long string background, with world-sheet fermionic matter fields of arbitrary representation, we can examine the interplay between integrability and nonlinear Poincare in more detail. 

When we have at least three bosonic flavors $(\text{i.e. } D>4)$, the presence of annihilations/reflections in the four-particle scattering is at odds with the Yang-Baxter equation (see, e.g. \cite{dorey} for an introduction) for the six-particle scattering S-matrix \cite{ali2}. For two flavors $(\text{i.e. } D=4)$ such four-particle annihilations/reflections do not necessarily preclude integrability, however it was shown that the Yang-Baxter equation is violated by calculating the one loop six particle scattering. Now including world-sheet massless fermions introduces two questions: For $D=4$, is there a fermionic matter field content that modifies the six particle S-matrix in such a way to restore the Yang-Baxter equation, and thus integrability? and for $D>4$, how does including massless fermions change the condition for the absence of bosonic annihilations/reflections? Here we shall examine the second question and find the relationship between the critical bulk dimension and the fermionic matter field content imposed by integrability.

\subsection{Bosonic String}
\begin{figure}[H]
\includegraphics[scale=0.08]{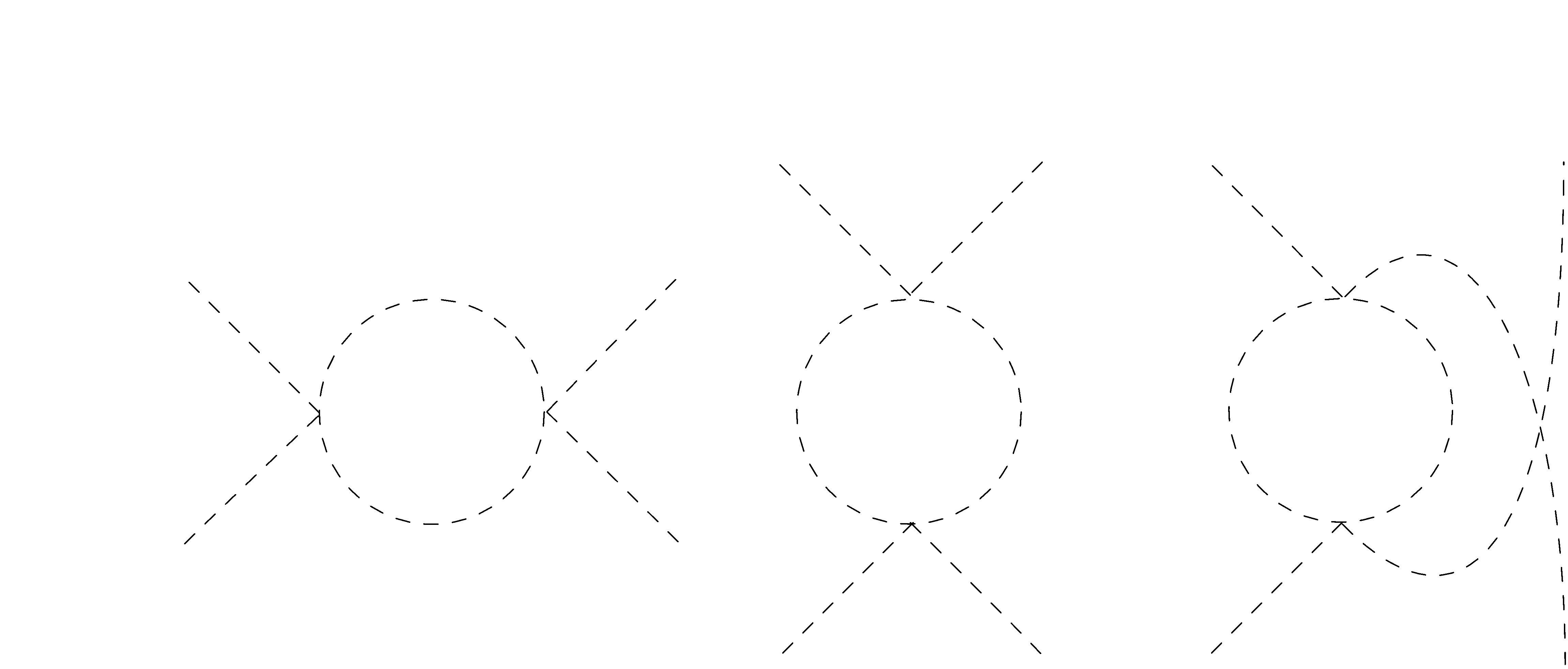}
\caption{The bosonic one loop contribution to the 4 boson annihilation. ($s$, $t$ and $u$ channels)}
\label{bosonicloop}
\end{figure}

To answer this question, let's review the situation for purely bosonic strings. In the absence of any additional matter fields, the Lagrangian given by the coset construction is the Nambu-Goto Lagrangian in addition to higher derivative geometric invariants such as the leading Polyakov-Kleinert term \cite{polyakov,kleinert}
\begin{equation}
\mathcal{L} =  \text{det }e\left( 1 + \alpha \Phi^i_{a\alpha}\Phi^a_{i\beta}e^{\alpha c}e^\beta_c + \ldots\right)
\end{equation}
The vanishing of annihilations in the process $X_i(p_1) X_i(p_2) \rightarrow X_j(p_3) X_j(p_4)$ ($i\neq j$) at tree level is guaranteed by the coefficients of the Nambu-Goto action which are fixed by nonlinearly realizes symmetry. However it was shown \cite{Sergei} that at one loop (fig \ref{bosonicloop}) we have a finite contribution to the annihilations given by
\begin{align}
\begin{split}
\mathcal{A} ^{\text{boson}}_{\text{finite}}&=
\frac{\ell_s^4}{192\pi} 
\Big( 
\left(26 - D\right) s^3 - stu \left( \frac{16}{3}D + \frac{4}{3} - 2 \left( D - 8\right) \text{log }\frac{-s}{\mu^2}\right) 
\\&\qquad\qquad\qquad- 12 t u \left( t \text{log } \frac{ s}{t} + u \text{log }\frac{s }{u} \right) 
\Big)
\end{split}\label{finiteB}
\end{align}
where the Mandlestam variables are the conventional  
\begin{equation*}
s = -(p_1 + p_2)^2 \qquad t = -(p_1 - p_3)^2 \qquad u = - (p_1 - p_4)^2
\end{equation*}
For two dimensional kinematics we will choose $t = 0 \Rightarrow u = -s$, which shows how annihilations only vanish for the critical dimension $D=26$. Away from the critical dimension, this loop provides a nonzero contribution to the flavor changing annihilation process, thus destroying integrability and reproducing the Polchinski-Strominger \cite{PS} result in the static gauge.

\subsection{Including Fermions}
Now we include $N$ $SO(2)$ Majorana-Weyl fermions with an arbitrary representation $R$ of $O(D-2)$ so that $J^{(R2)}_{ij}$ are the generators of $O(D-2)$ rotations. For our purposes we need to consider the most general weakly coupled Lagrangian up to quartic order in fields, so we need to expand the following in bosonic fields
\begin{equation}\begin{split}
\mathcal{L} &= \text{det }e\bigg(1 + i\,e^\alpha_a \bar{\theta} \rho^a \nabla_\alpha\theta \\
&\qquad
+ \sum_A \xi_A e^\alpha_a e^\beta_b\bar{\theta}\rho^a (1+\alpha_A \rho^*)\Gamma^A  \nabla_\alpha \theta\,\, \bar{\theta}\rho^b(1+\beta_A\rho^*)\Gamma_A  \nabla_\beta \theta \\
&\qquad\qquad
+ \sum_A \eta_A e^\alpha_b e^\beta_a\bar{\theta}\rho^a (1+\alpha'_A \rho^*)\Gamma^A  \nabla_\alpha \theta\,\, \bar{\theta}\rho^b(1+\beta'_A\rho^*)\Gamma_A  \nabla_\beta \theta + \ldots\bigg)
\end{split}\end{equation}
where $\xi, \eta, \alpha, \alpha', \beta, \beta'$ are arbitrary constants, and depending on the $SO(D-2)$ representation
\begin{equation}\begin{split}
& \Gamma^A_{ab} = \delta_{ab}  \qquad\qquad\qquad\qquad\qquad\qquad\,\, \text{scalar representation} \\
& \Gamma^A _{ab}= \mathds{1}_{ab},\gamma^i_{ab},\gamma^{[ij]}_{ab},\gamma^{[ijk]}_{ab},\ldots \qquad\qquad \text{spinor representation}\\
& \Gamma^A_{ab} = \delta_{ab},\delta^c_a \delta^d_b \qquad\qquad\qquad\qquad\qquad \text{vector representation}
\end{split}\end{equation}
Expanding in terms of bosons we can substitute $e^\alpha_a = \delta^\alpha_a + e^{(2)\alpha}_a +\ldots$ and \\$\nabla_\alpha = \partial_\alpha + iJ_{ij}^{(R2)}\mathcal{V}^{(2)ij}_\alpha + \ldots$ in the Lagrangian to obtain

\begin{align}
\mathcal{L} =& i \,\text{det }e \left(1+i\bar{\theta}\slashed{\partial}\theta\right) + i e^{(2)\alpha}_a \bar{\theta}\rho^a\partial_\alpha\theta-\bar{\theta} J_{ij}^{(R2)} \slashed{\mathcal{V}}^{(2)ij}\theta \nonumber\\
&\qquad
+ \sum_A \xi_A\bar{\theta} (1+\alpha_A \rho^*)\Gamma^A  \slashed{\partial} \theta\,\, \bar{\theta}(1+\beta_A\rho^*)\Gamma_A  \slashed{\partial} \theta \label{lagg}\\
&\qquad\qquad
+ \sum_A \eta_A\bar{\theta}\rho^a (1+\alpha'_A \rho^*)\Gamma^A  \partial_\alpha \theta\,\, \bar{\theta}\rho^a(1+\beta'_A\rho^*)\Gamma_A  \partial_\alpha \theta + \ldots \nonumber
\end{align}
as explained in the following subsection, we will only be interested in the quadratic fermionic vertices. So we consider the first line of eq(\ref{lagg})  and expand $e$ in eq(\ref{ff}) and $\mathcal{V}$ in eq(\ref{UV2}) in powers of derivatives to obtain the interaction terms between two bosons and two fermions
\begin{align}
\mathcal{L} &= \partial_\alpha X_i \partial^\alpha X^i	-i \bar{\theta} \slashed{\partial} \theta
 +\sum_{A=1}^{N} \ell_s^2 \bar{\theta}^A  \frac{1\pm\rho^*}{2} \left( i \rho_c \partial_a X_i \partial_b X^i \left(\eta ^{a b} \overset{\leftrightarrow}{\partial}\left.\right.^c -  \eta^{c(a}\overset{\leftrightarrow}{\partial}\left.\right.^{b)} \right) \right)\theta^A \nonumber \\
	&\qquad
	+\sum_{A=1}^{N} \ell_s^2 \bar{\theta}^A  \frac{1\pm\rho^*}{2}\left( \tfrac{i}{2} \sigma J^{(R2)}_{ij} \partial X^{[i} \slashed{\partial} \cdot \partial X^{j]} \right)\theta^A  	+ 
	\mathcal{O}\left( \left(\bar{\theta}\partial\theta\right)^2, \left(\partial X\right)^4\right)  \label{laggg}
\end{align}
Where $\pm$ is dependent on the chirality of each of our $N$ fermions, and $\sigma=0,1$ for scalar and non-scalar under $SO(D-2)$ respectively.  

\subsection{Tree Level Integrability}
eq(\ref{lagg}) clearly shows the huge freedom in choosing the four fermion vertices. These coefficients might be relevant for two loop integrability, and they are crucial for supersymmetric models which exhibit tree level fermionic flavor reflections and transmissions. More importantly, in addition to requiring the absence of tree level four fermion annihilations, integrability requires that all quintic processes vanish. Indeed because of the special properties of gamma matrices in each dimension, quintic tree level processes, particularly those invovling four fermions and one boson, were shown to vanish for supersymmetric models only for the critical number of dimensions \cite{ali2}. However in our case the freedom in the choice of the four fermion vertices renders tree level integrability a weak criterion, because such vertices directly affect the quintic processes. The one loop two-boson two-fermion process is also not a strong criterion now because of this freedom. 

\subsection{One Loop Integrability}
Fortunately the two-boson two-fermion vertices arise from the fermionic kinetic term, therefore the fermionic contribution to the one loop four boson annihilation process can be used as a model independent criterion for integrability for theories with nonlinearly realized poincare symmetry and additional massless fermionic matter fields.

\begin{figure}[H]
\includegraphics[scale=0.08]{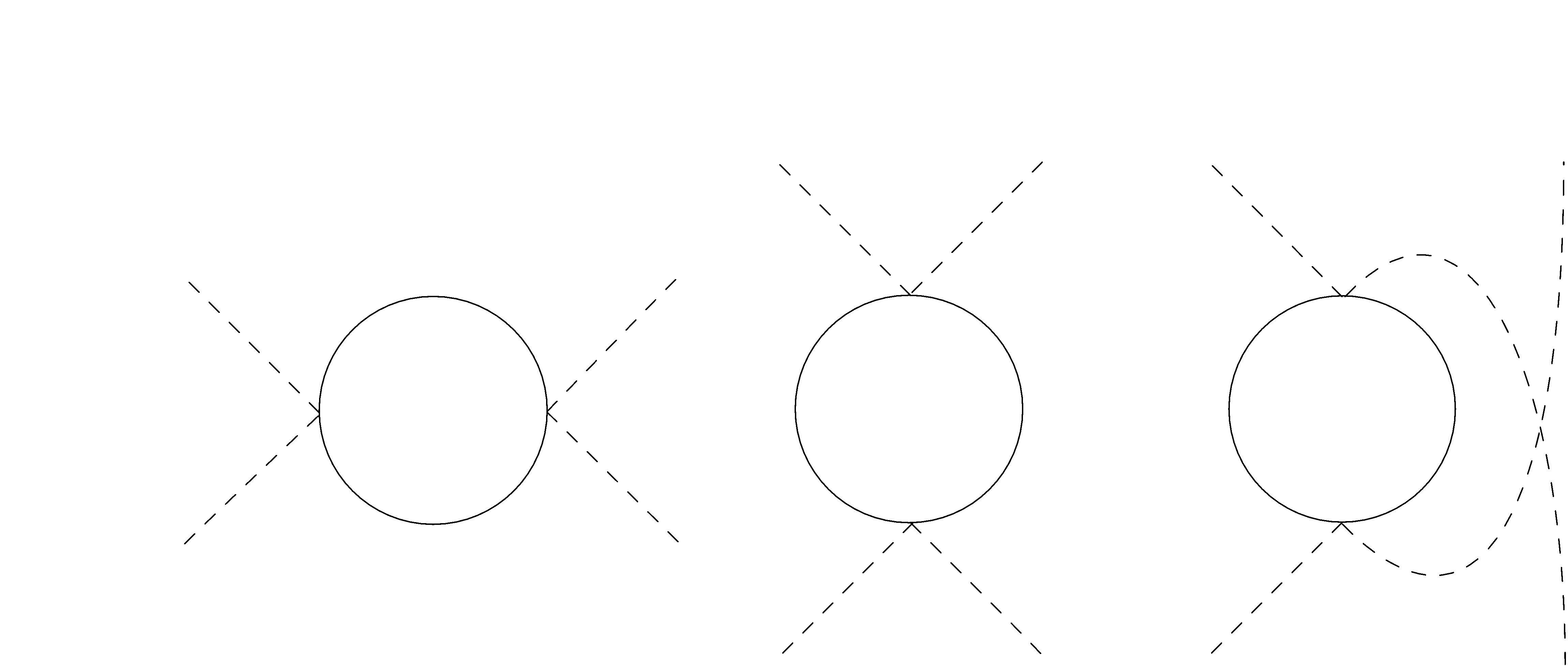}
\caption{The fermionic one loop contribution to the 4 boson annihilation. ($s$, $t$ and $u$ channels)}
\label{fermionicloop}
\end{figure}

The two-boson two-fermion vertices in eq(\ref{lagg}) give an infinite and a finite fermionic loop contribution (fig \ref{fermionicloop}) to the four boson annihilation scattering 
\begin{align}
\mathcal{M}_{\text{infinite}} &= -\ell_s^4\frac{ N}{2}\frac{ \text{dim}(R)}{192 \pi} \left( \frac{1}{\epsilon} + \gamma - \text{log }4\pi \right) st u \\
\mathcal{M}_{\text{finite}}  &= \ell_s^4 \frac{N}{4}\frac{\text{dim}(R)}{192 \pi} \left( s^3\left(1 + 3\sigma\right)  +  s t u \left( \frac{4}{3}-3\sigma - 2 \log\frac{-s}{\mu^2}\right)   \right) \label{finiteF}
\end{align}

Now we add the finite bosinic eq(\ref{finiteB}) and fermionic eq(\ref{finiteF}) contributions to the loop, and we find that the condition for the critical dimension from one-loop integrability becomes
\begin{equation}
\frac{1}{192\pi} \left( D - 26 + \frac{N}{4}\mathrm{dim }R \left(1 + 3 \sigma \right)\right)\label{oneloopintegrability}
\end{equation}

Now we can check for which dimensions $D$, number of fermions $N$ and representations $R$ under $SO(D-2)$ does this contribution vanish. For the vector representation the dimension is $2(D-2)$, whereas for spinors we use the fundamental representation in each dimension, which  (depending on whether we have Majorana and/or Weyl conditions in each dimension) is either $\frac{1}{2}$ or $\frac{1}{4}$ of the Dirac representation  $2^{[D-2]/2}$. For the scalar representation ($\sigma=0$ and $\mathrm{dim }R=1$), consistent with the Heterotic superstring and fermionization on the world sheet we see that eq(\ref{oneloopintegrability}) vanishes for any dimension if we choose $N=4 (26-D)$ Majorana Weyl fermions. The result for all representations is tabulated below\footnote{For the reducible spin 3/2 and symmetric representation, as well as the irreducible antisymmetric and traceless symmetric representations under $SO(D-2)$ we don't find any instance where eq(\ref{oneloopintegrability}) vanishes, except for $D=4$ and $N=11$, the antisymmetric representation which is equivalent to the Majorana and Vector representation for $SO(2)$} (Table \ref{table:nonlin})

confirming the well known result that $D=10$ is the critical dimension for supersymmetric strings with a single Majorana spinor from the Green Schwartz point of view, or a single $O(D-2)$ vector from the Ramond-Neveau-Schwarz point of view. Notice however that supersymmetry was not required a priori at this level to ensure integrability and that nonlinear Lorentz invariance alone at this order produces the integrable Lagrangian eq(\ref{laggg}), which is nothing but the gauge fixed Green-Schwarz Lagrangian.

\begin{table}[H]
\caption{One Loop Integrability} 
\centering 
\begin{tabular}{c c c c c c} 
\hline\hline 
Superstring &  $SO(D-2)$ Repr. & dim Repr. & $N$umber & $D$imension  \\ [0.5ex] 

\hline 
--- 					& Vector			&12  			& 1 			& 14			\\
\textbf{RNS}			& \textbf{Vector}	&\textbf{8}  	& \textbf{2} 	& \textbf{10}	\\
--- 					& Vector			&6  			& 3 			& 8			\\
--- 					& Vector			&3  			& 7 			& 5			\\
\textbf{GS}		 	& \textbf{Maj. Weyl}	&\textbf{8} 	& \textbf{2} 	& \textbf{10}	\\
---					& Weyl, Vector	 	&4			& 5			& 6			\\
---					& Majorana, Vector 	&2			& 11			& 4			\\ 	
\textbf{Heterotic} 		& \textbf{Scalar	}	&\textbf{1}   	& \textbf{64}	& \textbf{10}	\\
--- 					& Scalar			&1   			& 4(26-$D$)	& $0<D<26$	\\[1ex] 
\hline 
\end{tabular}
\label{table:nonlin} 
\end{table}

\begin{figure}[H]
\centering
\includegraphics[scale=0.08]{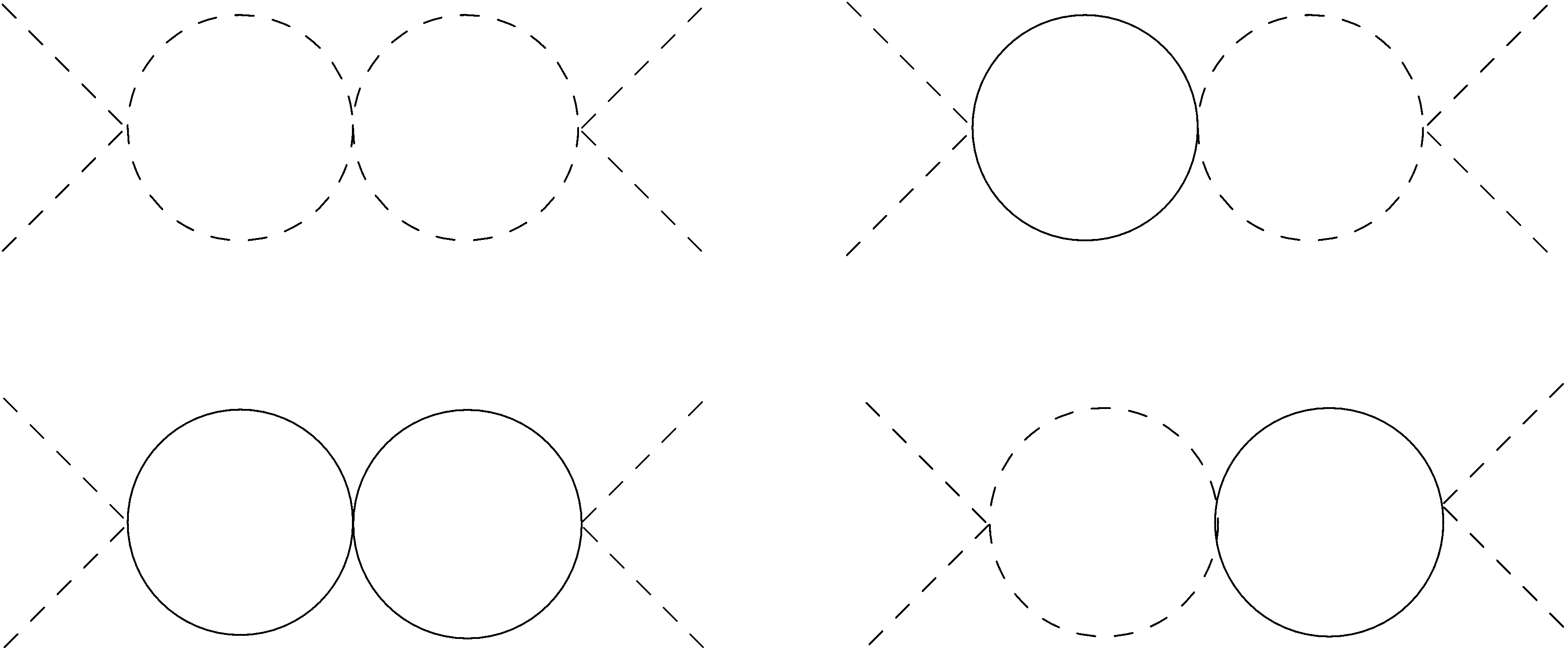}
\caption{Two loop s channel contribution to annihilations.}
\label{stwoloop}
\end{figure}

In all these cases the Polchinski-Strominger term \cite{PS} vanishes ensuring integrability to first loop order where supersymmetry is not necessarily present. Whether this remains true for higher loop orders is unclear. Even though a rigorous proof for higher loops requires knowing the coefficients of higher order vertices such as the four fermion vertex (Fig \ref{stwoloop}), one can still argue that anomalies usually manifest at one loop, and if they don't then it is likely that higher derivative counter terms are always sufficient to absorb the finite contribution in a way that preserves our nonlinear symmetries. 

Tree level integrability on the other hand is demonstrably restrictive, as shown in \cite{ali2} for quintic processes involving four fermion and one boson in the GS action for $N>1$. In that case supersymmetry and tree level integrability fixed the coefficients of the four fermion vertices directly, but the target space dimension was only restricted to be $3,4,6$ or $10$. Our result here eliminates the first three possible dimensions and singles out $D=10$, that is if we impose supersymmetry.

\begin{figure}[H]
\centering
\includegraphics[scale=0.08]{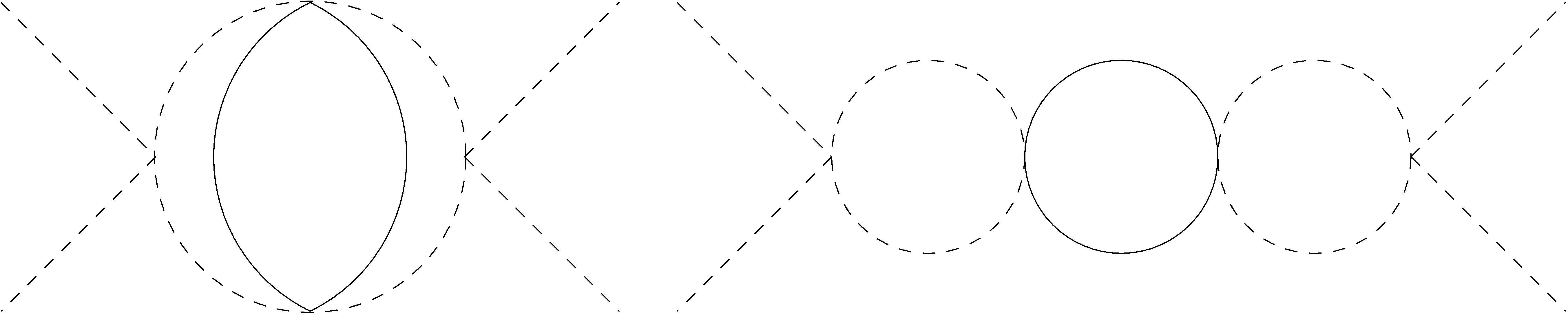}
\caption{Higher loop diagrams that will give rise to $C_2(R)$}
\label{threeloop}
\end{figure}

It is interesting to notice that at even higher loop order, we will have for some diagrams such as in (Fig \ref{threeloop}) fermion vertices with bosons of different flavor where these flavors are summed over, this will give rise to the quadratic Casimir of the fermions' representation under $SO(D-2)$. This can (and should) still however be consistent with both Green Schwarz and RNS superstrings, particularly because of spin triality. For vector representations $C_2(R) = D-3$, whereas for spinorial representations $C_2(R) = (D-2)(D-3)/8$, these two are only equal for $D=10$.

\section{Conclusions}
The detailed results of this paper elucidate the restrictions imposed by nonlinearly realized Poincare symmetry on the interactions between fermions of arbitrary representations and Goldstone bosons on the world sheet of strings in flat background. This paves the way for investigating the role played in integrability, by hidden symmetries of the Poincare group which is neither semi-simple nor an example of a SSM (symmetric space model)\cite{polyakov2}.

Our application here was to consider how much does one loop integrability alone restrict the class of Lorentz invariant theories. It turned out that there was still a certain level of freedom, and it seemed that analyzing tree level processes for the big variety of coefficients, might completely fix our integrable theory by eliminating the $D\neq 10$ possibilities in Table \ref{table:nonlin}. That would provide a novel way to re-derive the supersymmetric string. If on the other hand the coefficients were not fixed, that would also give rise to the exciting possibility of a new integrable Lorentz invariant string, with additional massless non-supersymmetric fermions.

\paragraph{Acknowledgements\\}
I would like to thank Sergei Dubovsky, Victor Gorbenko and Patrick Cooper for helpful discussions and useful insight.

\appendix

\section{Using BCH and Projecting out}\label{appendixA}
We use the Baker-Campbell-Haussdorff Lemma in two forms
\begin{align}
e^{-X}\delta e^{+X} 	&=	\delta X - \frac{1}{2!} [X,\delta X] + \frac{1}{3!} \big[ X, [X,\delta X] \big] - \frac{1}{4!} \Big[ X, \big[ X, [X, \delta X] \big] \Big]+ \ldots \label{BCH1}	\\
e^{-X}Y e^{+X}		&= Y - [X,Y] + \frac{1}{2!} \big[ X, [X,Y] \big] - \frac{1}{3!} \Big[ X, \big[X, [X, Y]\big] \Big] + \ldots\label{BCH2}
\end{align}
For example, we use the first in expanding the Cartan form $g^{-1} d g$ where $g = e^{i\sigma^{a}P_{a}+iX_{i}(\sigma)P_{i}}e^{i\phi_{ai}(\sigma)J_{ai}}$
\begin{equation}\begin{split}
g^{-1}\partial_\alpha g 		&= e^{-i\phi_{ai}J_{ai}} e^{-i\sigma^{a}P_{a}-iX_{i}P_{i}} \partial_\alpha \left( e^{i\sigma^{a}P_{a}+iX_{i}P_{i}}e^{i\phi_{ai}J_{ai}}\right) \\
						&= e^{-i\phi_{ai}J_{ai}}\left( e^{-i\sigma^{a}P_{a}-iX_{i}P_{i}} \partial_\alpha  e^{i\sigma^{a}P_{a}+iX_{i}P_{i}}\right)e^{i\phi_{ai}J_{ai}} 
							+e^{-i\phi_{ai}J_{ai}}\partial_\alpha e^{i\phi_{ai}J_{ai}} \\
						&= e^{-i\phi_{ai}J_{ai}}\left( iP_\alpha + i\partial_\alpha X_i P_i\right)e^{i\phi_{ai}J_{ai}} 
							+e^{-i\phi_{ai}J_{ai}}\partial_\alpha e^{i\phi_{ai}J_{ai}} 	
\end{split}\end{equation}
Now we apply eq(\ref{BCH1}) to each of the three terms individually. The first two will give the expected transformation of a vector under Lorentz transformations
\begin{align*}
e^{-i\phi_{ai}J_{ai}} P_je^{i\phi_{ai}J_{ai}} &=P_a \phi^{ia} \left( \frac{\text{sin }\sqrt{\phi\phi^T}}{\sqrt{\phi\phi^T}}  \right)_{ij} 
			+P^i \left(\text{cos }\sqrt{\phi\phi^T}\right)_{ij}\\
e^{-i\phi_{ai}J_{ai}}P_\alpha e^{i\phi_{ai}J_{ai}}		&= P_\alpha +P^a \phi^i_a\phi^j_\alpha \left( \frac{\text{cos }\sqrt{\phi\phi^T}-\mathds{1}}{\phi\phi^T}  \right)_{ij}
			+P^j  \phi^i_\alpha \left( \frac{\text{sin }\sqrt{\phi\phi^T}}{\sqrt{\phi\phi^T}}  \right)_{ij}
\end{align*}
These give us the frame field eq(\ref{framefield}) and the Goldstone derivative eq(\ref{implicit}), whereas for the third we use eq(\ref{BCH2}) to derive the recursion relations eq(\ref{UV1}) and eq(\ref{UV2})
\begin{align*}
e^{-i\phi_{ai}J_{ai}}\partial_\alpha e^{i\phi_{ai}J_{ai}} &=i\mathcal{V}^{ij}_\alpha(\phi)J_{ij}+\,i\mathcal{U}_\alpha(\phi)J_{+-}+\Phi_{\alpha,ai}(\phi)J_{ai}
\end{align*}
The recursion for $\Phi_{\alpha, ai}(\phi)$ was not provided but can be deduced from the same equation.

Now we turn to inverting
\begin{equation}
 \partial_{\alpha}X_{i} \left(\text{cos }\sqrt{\phi\phi^T}\right)^{ij} + \left( \phi \left( \frac{\text{sin }\sqrt{\phi\phi^T}}{\sqrt{\phi\phi^T}}  \right)\right)^{j}_{\alpha}=0\label{ffff}
\end{equation}
to solve for $\phi$ in terms of $X$ in the equation
\begin{equation}
 \partial_{\alpha}X_{i} =-  \phi_\alpha^j \left( \frac{\text{tan }\sqrt{\phi\phi^T}}{\sqrt{\phi\phi^T}}  \right)_{ij}
\end{equation}
we first calculate 
\begin{align*}
\partial_\alpha X_i \partial^\alpha X_j&= \phi_\alpha^{j'} \left( \frac{\text{tan }\sqrt{\phi\phi^T}}{\sqrt{\phi\phi^T}}  \right)_{ij'}\phi^{\alpha i'} \left( \frac{\text{tan }\sqrt{\phi\phi^T}}{\sqrt{\phi\phi^T}}  \right)_{i'j} \\
&=\left(\text{tan}^2\sqrt{\phi\phi^T}\right)_{ij}
\end{align*}
which allows us to find $\sqrt{\phi\phi^T}$ in terms of $\left(\sqrt{\partial_\alpha X \partial^\alpha X^T }\right)_{ij} $. Then it is straight forward to solve eq({\ref{ffff})

\section{Expanding in Powers of Fermions}\label{appendixB}

In this appendix we perform a functional form of the Taylor expansion in order to extract the fermionic fields from the square root in Volkov-Akulov action and the Green-Schwarz action. This is important if we wan to explicitly compare the terms in this series with the action constructed by adding fermionic terms covariantly as matter fields using the CCWZ procedure. This is also especially helpful if we want to calculate low energy scattering amplitudes with a specific number of fermionic states, or if we were interested in low dimensional theories where the expansion terminates quickly due to the anti-commuting statistics of the fermion field components.

\begin{align}
\mathcal{L}_{AV} 	&= \sqrt{\text{det }\Pi_\alpha \cdot \Pi_\beta} \nonumber\\
 		&=  \left. \mathcal{L}_{AV} \right\rvert_{\psi=0}  + i\bar{\psi} \Gamma_\mu \partial_\beta \psi \int \left. \frac{\delta \mathcal{L}_{AV}}{\delta i\bar{\psi}\Gamma_\mu \partial_\beta\psi}\right\rvert_{\psi=0} + h.c. +\ldots
\end{align}

and
\begin{align}
\left. \frac{\delta \mathcal{L}_{AV}}{\delta i\bar{\psi} \Gamma_\mu \partial_\beta \psi}\right\rvert_{\psi=0} &= \frac{1}{\mathcal{L}_{AV}} \text{Tr }\left.\left(\text{adj }(\Pi\cdot\Pi)\frac{\delta\Pi\cdot\Pi }{\delta i\bar{\psi} \Gamma_\mu \partial_\beta \psi}\right)\right\rvert_{\psi=0} \nonumber \\
 &= \frac{1}{\mathcal{L}_{AV}} \left.\text{adj }(\Pi\cdot\Pi)^{\alpha\delta}\Pi_{(\delta}^\nu\frac{\delta\Pi_{\alpha)\nu} }{\delta i\bar{\psi} \Gamma_\mu \partial_\beta \psi}\right\rvert_{\psi=0} \nonumber \\
 &= \frac{2}{\mathcal{L}_{AV}} \left.\text{adj }(\Pi\cdot\Pi)^{\alpha\beta}\Pi_\alpha^\mu\right\rvert_{\psi=0} \nonumber\\
 &= \frac{2}{\mathcal{L}_{NG}} \left(  \partial^\beta X^\mu \partial_\alpha X^\nu \partial^\alpha X_\nu - \partial^\beta X_\nu\partial_\alpha X^\nu \partial_\alpha X^\mu  \right) 
\end{align}

Here $\mathcal{L}_{NG}\equiv\mathcal{L}_{AV}|_{\psi=0}$. 

Now for $D=3$ and substituting $X^\alpha = \sigma^\alpha$ we get
\begin{align}
\left. \frac{\delta \mathcal{L}_{AV}}{\delta i\bar{\psi} \rho_\gamma \partial_\beta \psi}\right\rvert_{\psi=0} &= 
\frac{2}{\mathcal{L}_{NG}} \left(  \eta^{\beta\gamma} \partial_\alpha X^\nu \partial^\alpha X_\nu - \partial^\beta X_\nu\partial^\gamma X^\nu   \right)\nonumber \\
\left. \frac{\delta \mathcal{L}_{AV}}{\delta i\bar{\psi} \rho^* \partial_\beta \psi}\right\rvert_{\psi=0} &= 
\frac{2}{\mathcal{L}_{NG}} \left(  \partial^{[\beta} X \partial^{\alpha]} X^\nu \partial_\alpha X_\nu  \right)
\end{align}
So that
\begin{align}\label{VAExpand}
\mathcal{L}_{AV} 	=  \mathcal{L}_{NG}    +& \frac{2i\bar{\psi} \rho_\gamma \partial_\beta \psi}{\mathcal{L}_{NG}} \left(  \eta^{\beta\gamma} (1 + (\partial X)^2) - \partial^\beta X\partial^\gamma X   \right) \nonumber\\  &+ \frac{2i\bar{\psi} \rho^* \partial_\beta \psi}{\mathcal{L}_{NG}} \partial^\beta X + h.c. +\mathcal{O}\left( (\bar{\psi}\rho\partial\psi)^2\right)
\end{align}

\section{Special case of $D=3$}\label{appendixC}
For $D=3$ we only have one boson field $X$, so the frame field, goldstone field derivative and gauge functions in eq(\ref{transformations2}) simplify to
\begin{align}
e_{a \alpha} &= \eta_{a \alpha} + \varphi_a \varphi_\alpha\, \tfrac{ \text{cos}|\varphi| - 1} { \varphi^2} - \varphi_a \partial_\alpha X \,\text{sinc}|\varphi| \label{oneone} \\
D_\alpha &= \partial_\alpha X \text{cos}|\varphi|  + \varphi_\alpha \text{sinc}|\varphi|\label{twotwo}\\
U_\alpha &= \epsilon^{ab} \varphi_a \partial_\alpha \varphi_b \,\,\tfrac{ \text{cos} |\varphi| - 1 }{\varphi^2}\label{threethree}\\
\Phi_{a \alpha} &= \partial_\alpha \varphi_a + \varphi_a \epsilon^{bc} \varphi_b \partial_\alpha \varphi_c \,\,\tfrac{ \text{sinc} |\varphi| - 1}{|\varphi|^2}\label{fourfour}
\end{align}
similarly we can explicitly derive the transformation rules for $\sigma, X$ and $\varphi_\alpha$ using eq(\ref{transformation}) which gives

\begin{align*}
P\left( \delta ^a_\epsilon X + \epsilon \sigma^a \right) + P_b \left( \delta^a_\epsilon \sigma^b + \epsilon \eta^{ab} X\right) &= 0 \\
e^{-\varphi\cdot J} \delta_\epsilon ^a e^{+\varphi\cdot J}   -\epsilon\, e^{-\varphi\cdot J} J^a e^{+\varphi\cdot J} + u_\epsilon B &= 0
\end{align*}
The first of these two equations gives us the transformation of the physical field $X'(\sigma') = X(\sigma) + \epsilon \sigma^a$ and of the world sheet coordinate $\sigma'^b = \sigma^b + \epsilon \eta^{ab} X(\sigma)$, whereas the second equation expands to
 \begin{align*}
e^{-\varphi\cdot J} \left(\delta^a-J^a\right) e^{+\varphi \cdot J} &=
B\epsilon^{bc} \left( \varphi_b \delta^a \varphi_c \tfrac{ \text{cos}|\varphi| - 1}{\varphi^2} -  \varphi_b \delta^a_c\, \text{sinc} |\varphi| \right) + B u_\epsilon
\\&\qquad\qquad+ 
J^b \bigg( -\varphi_b \left(\varphi\cdot \delta^a \varphi\right) \tfrac{\text{sinc}|\varphi| - 1}{\varphi^2} +\delta^a \varphi_b \,\text{sinc}|\varphi|
\\&\qquad\qquad\qquad\qquad\qquad
 -\delta_b^a \,\text{cos}|\varphi|  + \varphi^a \varphi_b \tfrac{\text{cos}|\varphi| - 1}{\varphi^2}\bigg)
\end{align*}
and gives us the transformation of the auxiliary field as well as the function $u$ appearing in the transformation of the covariant matter fields 
\begin{align}
\delta^a \varphi_b &= \frac{ |\varphi| }{\text{tan}|\varphi|} \left( \delta ^a_b - \frac{\varphi_b \varphi^a}{\varphi^2}\right) + \frac{\varphi_b \varphi^a}{\varphi^2} \\
u\left(\varphi\right) &= \frac{ \epsilon^{ba}\varphi_b}{|\varphi|} \left( \text{sin}|\varphi| - \frac{ \text{cos}|\varphi| - 1}{ \text{tan}|\varphi|}\right)\label{ud3}
\end{align}

\paragraph{Simplest Action\\}
Now we can  write down the Lagrangian quadratic in spinor matter fields 
\begin{align*}
\mathcal{L} &= \text{det }e \left( 1 + \alpha_1 \bar{\psi} \rho^a e_a^{\alpha} \left( \partial _\alpha - \rho^* U_\alpha\right)\psi + \alpha_2 \bar{\psi} \rho^a \rho^b \psi e_a^\alpha \Phi_{\alpha b} \right)
\end{align*}
where eq(\ref{oneone}) gives the determinant and the inverse as 
\begin{align*}
\text{det }e &= - \text{cos}|\varphi| + \left( \varphi \cdot \partial X\right) \tfrac{\text{sin}|\varphi|}{|\varphi|}
\\
e^\alpha_a &= \delta^\alpha_a - \frac{ \varphi^\alpha\varphi_a \tfrac{\text{cos }\varphi - 1}{\varphi^2}  - \varphi^\alpha \partial_a X \text{sinc }\varphi    } {\text{cos }\varphi - \varphi\cdot\partial X \text{sinc }\varphi}
\end{align*}

For complete spinors the field redefinition in eq(\ref{redefinition}) and canonically normalized $\psi\rightarrow e^{\varphi\cdot\rho\rho^*}\psi/\sqrt{\alpha_1}$ gives
\begin{align}
\bar{\psi} \rho^a e^\alpha_a \partial_\alpha \psi &\rightarrow   \bar{\psi} \left(e^{-\varphi\cdot\rho\rho*}\rho^a    e^{+\varphi\cdot\rho\rho*}  \right)e^\alpha_a	\partial_\alpha\psi 
	+  e^\alpha_a \bar{\psi} e^{-\varphi\cdot\rho\rho*}\rho^a    \left( \partial_\alpha e^{+\varphi\cdot\rho\rho*} e^{-\varphi\cdot\rho\rho*} \right) e^{+\varphi\cdot\rho\rho*}	\psi \nonumber\\
&=		 \left(	\eta^{ab} - \varphi^a \varphi^b \tfrac{\text{cos}|\varphi| - 1}{\varphi^2} \right) e^\alpha_a \bar{\psi}\rho_b \partial_\alpha\psi  +  \varphi^a \text{sinc}|\varphi| e^\alpha_a\bar{\psi}\rho^*\partial_\alpha\psi \nonumber\\
	&\qquad\qquad+  e^\alpha_a \bar{\psi} e^{-\varphi\cdot\rho\rho*}\rho^a    \left( \rho^b\rho^* \Phi_{b\alpha} + \rho^* U_\alpha \right) e^{+\varphi\cdot\rho\rho*}	\psi 
\nonumber
\end{align}
where we used $ \left( \partial_\alpha e^{+\varphi\cdot\rho\rho*}\right) e^{-\varphi\cdot\rho\rho*} =  \rho^b\rho^* \Phi_{b\alpha} + \rho^* U_\alpha$. 

Therefore the Lagrangian becomes
\begin{align}	
\mathcal{L}&\rightarrow \text{det }e \bigg[
1+ \left(	\eta^{ab} - \varphi^a \varphi^b \tfrac{\text{cos}|\varphi| - 1}{\varphi^2} \right) e^\alpha_a \bar{\psi}\rho_b \partial_\alpha\psi  +  \varphi^a \text{sinc}|\varphi| e^\alpha_a\bar{\psi}\rho^*\partial_\alpha\psi 
 \nonumber\\ &\qquad\qquad 	+  e^\alpha_a \bar{\psi} e^{-\varphi\cdot\rho\rho*}\rho^a    \left(1+\alpha_2'\right)\rho^b\rho^* \Phi_{b\alpha}  e^{+\varphi\cdot\rho\rho*}	\psi \bigg] \nonumber\\
 &\xrightarrow{\alpha_{2}=-1}
 \text{det }e \bigg[
1- \left(	\eta^{ab} - \varphi^a \varphi^b \tfrac{\text{cos}|\varphi| - 1}{\varphi^2} \right) e^\alpha_a \bar{\psi}\rho_b \partial_\alpha\psi  - \varphi^a \text{sinc}|\varphi| e^\alpha_a\bar{\psi}\rho^*\partial_\alpha\psi \bigg] \label{simplestaction}
\end{align}

\newpage

\addcontentsline{toc}{section}{References}
\bibliographystyle{utphys}
\bibliography{Fermions on the World  Sheet of Effective Strings via Coset Construction}

\end{document}